\documentclass[12pt]{article}
\usepackage{amssymb,amsmath,epsfig}
\allowdisplaybreaks

\begin{document}
\title{\bf Phase Transition and Thermal Fluctuations of Quintessential Kerr-Newman-AdS Black Hole}
\author{M. Sharif \thanks{msharif.math@pu.edu.pk} and Qanitah Ama-Tul-Mughani
\thanks{qanitah94@gmail.com}\\
Department of Mathematics, University of the Punjab,\\
Quaid-e-Azam Campus, Lahore-54590, Pakistan.}

\date{}
\maketitle
\begin{abstract}
This paper is devoted to analyzing the critical phenomenon and phase
transition of quintessential Kerr-Newman-anti-de Sitter black hole
in the framework of Maxwell equal-area law. For this purpose, we
first derive thermodynamic quantities such as Hawking temperature,
entropy and angular momentum in the context of extended phase space.
These quantities satisfy Smarr-Gibbs-Dehum relation in the presence
of quintessence matter. We then discuss the critical behavior of
thermodynamic quantities through two approaches, i.e., van der
Waals-like equation of state and Maxwell equal-area law. It is found
that the latter approach is more effective to analyze the critical
behavior of the complicated black holes. Using equal-area law, we
also study phase diagram in $T-S$ plane and find an isobar which
shows the coexistence region of two phases. We conclude that below
the critical temperature, black holes show a similar phase
transition as that of van der Waals fluid. Finally, we study the
effects of thermal fluctuations on the stability of this black hole.
\end{abstract}

\section{Introduction}

Cosmological evidences indicate that our universe is undergoing an
accelerated expansion due to some large negative pressure. The most
prevailed conjecture to discuss the evolution of the cosmos is an
anti-gravitational force named as dark energy (DE). Despite of
enormous astronomical observations, the essential features as well
as the origin of DE is still unclear and has become a source of
vivid debate. This can be expressed by different models such as the
cosmological constant $(\Lambda)$, quintessence energy, etc. The
homogeneous cosmological constant or vacuum energy has fixed value
in space i.e., $\Lambda\approx1.3\times10^{-56}$cm$^{-2}$ \cite{25c}
whereas the quintessence energy is inhomogeneous as well as
dynamical scalar field which can be characterized by the equation of
state (EoS) $w= \frac{P}{\rho}$, where $P$ and $\rho$ are the
pressure and energy density, respectively. It is believed that the
late-time evolution may be the consequence of cosmological constant
or quintessence matter which permeates throughout the universe.
Thus, if quintessence matter is spread out all over the spacetime,
it must cover the black hole (BH) surrounding which alters its
spacetime structure as well as asymptotic features of the
cosmological horizon.

Wei and Liu \cite{fa} studied the relation between pressure and
cosmological constant for charged spherically symmetric anti-de
Sitter (AdS) BH in higher-dimensional spacetime. They found an
analogy between BH and the van der Waals (vdW) liquid-gas system
which kept the foundation of BH thermodynamics in extended phase
space. The extended phase space is based on the fact that
thermodynamic pressure is identified with the cosmological constant
and its conjugate quantity to the volume of BH. This implies that
the conventional phase space is quite different from the extended
phase space where the extra "PdV" term is present which modifies the
Smarr relation as well as the first law of BH. In this phase space,
the whole BH system is mapped to the vdW fluid system consolidating
the analogy between small/large BH with the vdW liquid/gas phase
transitions. It is known that above the critical temperature, the
isothermal curves in the vdW system depict similar behavior to the
experimental results. However, below the critical temperature, there
exists an oscillating region which violates the condition of stable
equilibrium. Using Maxwell equal-area law, the oscillating part can
be replaced with an isobar which yields the correspondence with the
experimental results. After this breakthrough, several aspects of
charged AdS BH have been discussed, such as $P-V$ or $T-S$
criticality, first-order phase transition, etc \cite{51}-\cite{b}.

Gunasekaran et al. \cite{45} derived the critical thermodynamic
quantities of charged and rotating AdS BH in extended phase space
and found its analogy with the vdW liquid-gas system. Cheng et al.
\cite{52} investigated the critical behavior of Kerr-Newman-AdS BH
in extended phase space and numerically solved the critical points
for the vdW-like phase transition.  Wei and Liu \cite{46} examined
the phase transition of charged AdS black hole in Gauss-Bonnet
gravity. It is found that the charged AdS BH, in the presence of
quintessence, shows a small-large BH phase transition similar to the
liquid-gas phase transition of the vdW system. Li \cite{44} examined
the effect of DE on $P-V$ criticality of Reissner-Nordstrom (RN) AdS
BH and showed that quintessence matter does not affect the existence
of small/large BH phase transition. Guo \cite{48} used Maxwell
equal-area law to determine the phase transition of charged AdS BHs
in the presence of quintessence matter and concluded that BHs have
the same phase transition as that of the vdW system.

In quantum gravity, one of the important issues is the consideration
of statistical perturbations which modify the BH geometry. These
perturbations lead to thermal fluctuations that do not affect
thermodynamics of large BHs but have great implications on the BHs
whose size and temperature decrease and increase, respectively, due
to Hawking radiation. \cite{1}. Thus, thermal fluctuations play a
critical role in the small BHs thermodynamics due to sufficient
increase in temperature. Pourhassan et al. \cite{18} derived
logarithmic corrections to entropy around the equilibrium state and
investigated their influence on the thermodynamics of
higher-dimensional charged BHs. They also investigated the vdW fluid
duality as well as the validity of the first law of thermodynamics.
Upadhyay \cite{41'} discussed the effects of leading order
corrections on the stability of charged rotating AdS BHs. They found
that thermodynamic potentials satisfy the first law of BH
thermodynamics. Moreover, the effect of first-order corrections on
the phase transition of Kerr-Newman-AdS BH has been analyzed
\cite{21}. It is found that these corrections modify the
thermodynamic potentials and its phase transition. In the same
perspective, Zhang \cite{22a} analyzed the physical behavior of
thermodynamic potentials of RN AdS and Kerr-Newman BHs in the
presence of first-order corrections. Recently, Sharif and Akhtar
\cite{23a} studied quasi-normal modes and thermal fluctuations of
charged black hole with Weyl corrections.

This work aims to explore the effects of DE on critical behavior and
phase transition of charged rotating BH surrounded by the
quintessential field. We discuss thermodynamic properties including
angular velocity and Smarr relation. We also derive the exact
expression of critical quantities through two approaches and discuss
the phase transition of BH. The paper is assembled as follows. In
the next section, we elaborate the spacetime structure and calculate
its thermodynamics quantities. Further, the $P-V$ criticality of
quintessential Kerr-Newman BH is investigated in extended phase
space. In section \textbf{3}, we use Maxwell equal-area law in $T-S$
conjugate variables to study the conditions satisfied by phase
transition. Section \textbf{4} provides the general expression of
corrected entropy as well as explores the effects of thermal
fluctuations on the stability of considered BH. Finally, we compile
our results in the last section.

\section{Quintessential Kerr-Newman-AdS Black Hole}

Using Newman-Penrose formalism \cite{25f}, Xu and Wang \cite{26a}
derived the Kerr-Newman-AdS BH in the presence of quintessence field
whose line-element, in Boyer-Lindquist coordinates, reads
\begin{equation}\label{1}
ds^{2}=-\frac{\chi}{\Omega}\bigg[dt-\frac{a\sin^2\theta}{k}
d\phi\bigg]^2 +\frac{\Omega}{\chi}dr^{2}+\frac{\Omega
}{\tilde{P}}d\theta^{2}+\frac{\tilde{P}\sin^{2}\theta}{\Omega
}\bigg[a dt-\frac{(r^2+a^2)}{k}d\phi\bigg]^2,
\end{equation}
with
\begin{eqnarray}\label{2}
\chi&=&(r^2+a^2)(1+\frac{r^2}{l^2})-2mr+q^2-\alpha r^{1-3w}, \\
\label{3} \Omega&=&r^2+a^2\cos^2\theta, \quad k=1-\frac{a^2}{l^2},
\quad \tilde{P}=1-\frac{a^2}{l^2}\cos^2\theta .
\end{eqnarray}
Here $a$ corresponds to rotation parameter and $q$ is defined as
$q^2=q_e^2+q_m^2$, where $q_e$ and $q_m$ represent the electric and
magnetic charges, respectively. Moreover, $m$ is the mass of BH,
$l^2=-\frac{3}{\Lambda}$ determines the radius of AdS BH, $w$ is the
dimensionless state parameter with $-1<w<-\frac{1}{3}$ and $\alpha$
is the quintessence parameter which measures the intensity of
quintessential field around a BH, satisfying the following
inequality \cite{26a}
\begin{equation}
\alpha\leq\frac{2}{1-3w}8^{w}.
\end{equation}
It is analyzed that the above relation holds until the cosmological
horizon determined by quintessential DE exists. The line element
(\ref{1}) can be re-written as
\begin{equation}\label{5}
ds^2=-\mathcal{F}(r,\theta)dt^2+\frac{dr^2}{\mathcal{G}(r,\theta)}
+\Sigma(r,\theta)d\theta^2+K(r,\theta)d\phi^2-2H(r,\theta)dtd\phi,
\end{equation}
where
\begin{eqnarray}
\mathcal{F}(r,\theta)&=&\frac{\chi-a^2 \tilde{P} \sin ^2\theta
}{\Omega},\quad \mathcal{G}(r,\theta)=\frac{\chi}{\Omega},\quad
\Sigma(r,\theta)=\frac{\Omega}{\tilde{P}},\\K(r,\theta)&=&\frac{\tilde{P}
\sin ^2\theta \left(a^2+r^2\right)^2-\chi \left(a \sin ^2\theta
\right)^2}{k^2 \Omega },\\H(r,\theta)&=&\frac{a \tilde{P} \sin
^2\theta \left(a^2+r^2\right)-\chi a\sin ^2\theta}{k\Omega}.
\end{eqnarray}
The associated electromagnetic potential is given as
\begin{eqnarray}\nonumber
B=\frac{1}{\Omega}[-(-q_{e}r+q_{m}a\cos\theta)dt+\frac{1}{k}(a\sin
^2\theta q_{e} r+q_{m}(r^2+a^2)\cos\theta)d\phi].
\end{eqnarray}
For $\alpha=0$, the line element (\ref{1}) reduces to
Kerr-Newman-AdS BH while the Kerr-AdS BH solution can be obtained
for $\alpha=0=q$.

\subsection{Critical Phenomenon and Thermodynamical Structure
in Extended Phase Space}

Black hole as an interdisciplinary area provides a possible bridge
between classical general relativity and quantum theory of gravity.
Based on the pioneering work of Hawking and Bekenstein \cite{1a}, BH
thermodynamic quantities such as temperature and entropy can be
mapped on the the laws of ordinary thermodynamics which have opened
many interesting aspects of unification of gravity, quantum
mechanics and thermodynamics \cite{3a}-\cite{5a}. In this section,
we will discuss thermal properties as well as $P-V$ criticality of
the quintessential Kerr-Newman-AdS BH in an extended phase space.
Using the horizon condition $\chi(r_{+})=0$, the mass of the BH in
terms of horizon radius $r_{+}$ reads
\begin{equation}\label{9}
m=\frac{1}{2 r_+}(r_+ \left(\frac{r_+
\left(a^2+l^2+r_+^2\right)}{l^2}-\alpha r_+^{-3 w}\right)+a^2+q^2).
\end{equation}
The entropy in terms of horizon area is defined as
\begin{equation}\label{10}
S=\frac{A}{4}=\frac{\pi  \left(a^2+r_{+}^2\right)}{k}, \quad
\text{with} \quad
A=\int_{0}^{2\pi}\int_{0}^{\pi}\sqrt{g_{\theta\theta}
g_{\phi\phi}}|_{r=r_{+}}d\theta d\phi,
\end{equation}
which plays a significant role to study the thermodynamic evolution
of BH. It is observed that the entropy depends upon rotation and
horizon radius whereas the cosmological constant, electric charge
and quintessence parameter have no explicit effect on it.
\begin{figure}\center
\epsfig{file=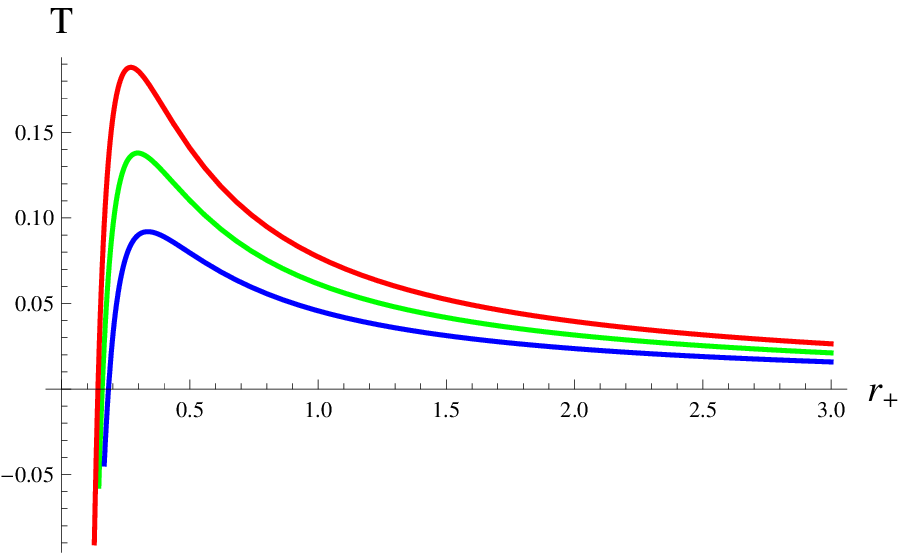,width=0.45\linewidth}
\epsfig{file=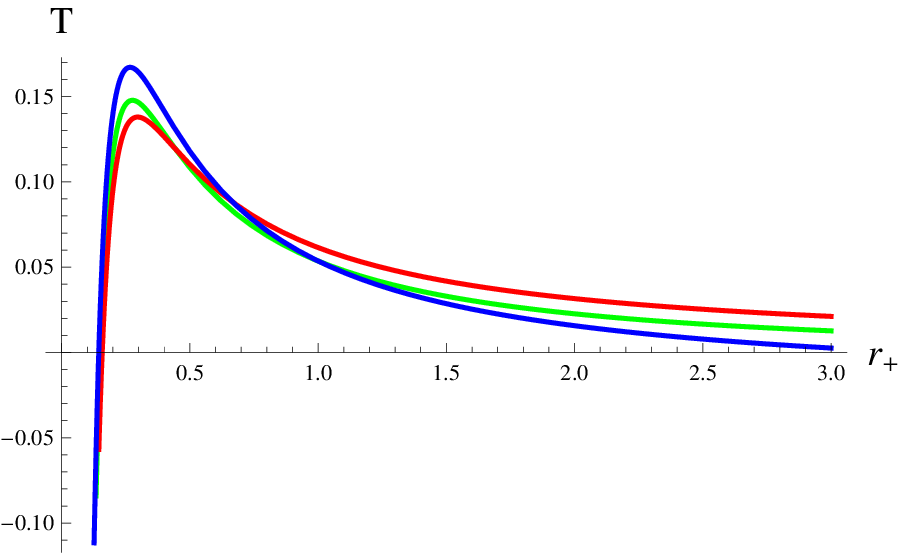,width=0.45\linewidth} \caption{Hawking
temperature vs $r_{+}$ for $a=0.1=q$. Left plot:  $w=\frac{-1}{3}$
and $\alpha=0$ (red), 0.2 (green ), 0.4 (blue). Right plot:
$\alpha=0.2$ and $w=\frac{-1}{3}$ (red), $w=\frac{-1}{2}$ (green),
$w=\frac{-2}{3}$ (blue).}
\end{figure}

The Hawking temperature
$(T=\frac{\chi^{'}(r)}{4\pi(r^2+a^2)}|_{r=r_{+}})$ for the
considered BH is evaluated as
\begin{eqnarray}\label{11}
T&=&\frac{1}{4 \pi \left(a^2+r_+^2\right)}\bigg(2 (\frac{r_+
\left(a^2+l^2+2 r_+^2\right)}{l^2}-m)+\alpha (3 w-1) r_+^{-3
w}\bigg).
\end{eqnarray}
Figure \textbf{1} shows the graphical behavior of Hawking
temperature with respect to horizon radius. It is observed that the
temperature of BH decreases for larger values of quintessence
parameter (left plot). For smaller values of state parameter, the
temperature increases and decreases, respectively, before and after
the critical radius (right plot). The corresponding angular velocity
is computed as
\begin{eqnarray}\label{13}
\Pi_{r}=-\frac{g_{t\phi}}{g_{\phi\phi}}=\frac{a P \sin ^2(\theta )
\left(a^2+r^2\right)-\chi a \sin ^2(\theta)}{P \sin ^2(\theta )
\left(a^2+r^2\right)^2-\chi \left(a \sin ^2(\theta)\right)^2}.
\end{eqnarray}
The radial function $\chi$ vanishes at the event horizon which
reduces the above expression to the following form
\begin{eqnarray}\label{14}
\Pi_{H}=\frac{a k}{\left(a^2+r_+^2\right)},
\end{eqnarray}
where $\Pi_{H}$ is the angular velocity of BH horizon. It is known
that the angular velocity of rotating BHs in AdS space does not
vanish at infinity, i.e., $\Pi_{\infty}\neq0$ for $r\rightarrow
\infty$ \cite{qa, ma}. This salient feature differentiates them from
the asymptotically flat spacetime where $\Pi_{\infty}=0$. For the
rotating BHs, the angular velocity is defined as
\begin{eqnarray}\label{15}
\Pi=\Pi_{H}-\Pi_{\infty}=\frac{a \left(l^2+r_+^2\right)}{l^2
\left(a^2+r_+^2\right)},
\end{eqnarray}
where the angular potential at infinity reads
\begin{eqnarray}\label{16}
\Pi_{\infty}=\frac{a }{l^2}.
\end{eqnarray}

In extended phase space, the cosmological constant is treated as
pressure and its conjugate quantity to the volume of BH \cite{45,
a}, given by
\begin{eqnarray}\label{17}
P&=&\frac{3}{8\pi l^2},
\end{eqnarray}
\begin{equation}\label{18}
V=\left(\frac{\partial M}{\partial P}\right)_{S,Q,J,\alpha}=\frac{2
\pi  \left(a^2 l^2 \left(q^2-\alpha r_+^{1-3
w}\right)+\left(a^2+r_+^2\right) \left(a^2 l^2-a^2 r_+^2+2 l^2
r_+^2\right)\right)}{3 k^2 l^2 r_+}.
\end{equation}
Here mass $M$, angular momentum $J$ and charge $Q$, are related to
parameters $m$, $a$ and $q$ as follows
\begin{eqnarray}
M=\frac{m}{k^2}, \quad J=\frac{ma}{k^2}, \quad Q=\frac{q}{k}.
\end{eqnarray}
Through Eqs.(\ref{10}) and (\ref{11}), we obtain
\begin{eqnarray}\nonumber
TS&=&\frac{1}{{4 k}}\bigg(-\frac{r_+^2 \left(8 \pi  a^2 P+3\right)+3
\left(a^2+q^2\right)+8 \pi Pr_+^4}{3r_+}\\\label{19}
&+&\frac{2}{3}r_+\left(8 \pi  a^2 p+16 \pi  P r_+^2+3\right)+\alpha
(3 w-1) r_+^{-3 w}+\alpha r_+^{-3 w}\bigg),
\end{eqnarray}
which in terms of physical parameters $M$, $J$ and $Q$ satisfies
Smarr-Gibbs-Dehum relation as
\begin{eqnarray}\label{20}
M&=&2(T_{k}S - PV + \Pi J ) + Q\Phi+\alpha  \Psi \left(-\frac{2
a^2}{a^2+r_+^2}+3 w+1\right),
\end{eqnarray}
where the electric potential $\Phi$ and physical quantity $\Psi$
(conjugate to the parameter $\alpha$) \cite{mm} read
\begin{eqnarray*}\nonumber
\Phi&=&\frac{q r_+} {a^2+r_+^2},  \quad \Psi=-\frac{r_+^{-3 w}}{2
k}.
\end{eqnarray*}
It is noted that for $\alpha=0$, all the derived quantities reduce
to charged rotating AdS BH \cite{45}.

Now we discuss the effects of quintessence matter on $P-V$
criticality of Kerr-Newman-AdS BH. Using Eqs.(\ref{9}) and
(\ref{17}), the Hawking temperature takes the form
\begin{equation}\label{24}
T=\frac{2 r_+ \left(8 \pi  P \left(a^2+2 r_+^2\right)+3\right)-r_+
\left(8 \pi  a^2 P+3\right)-\frac{3 \left(a^2+q^2\right)}{r_+}-8 \pi
P r_+^3+9 \alpha  w r_+^{-3 w}}{12 \pi  \left(a^2+r_+^2\right)}.
\end{equation}
In terms of charge and angular momentum, the EoS can be written as
\begin{eqnarray}\nonumber
P&=&\frac{T}{2 r_+}+\frac{Q^2}{8 \pi r_+^4}-\frac{1}{8 \pi
r_+^2}-\frac{3 \alpha  w r_+^{-3 w-3}}{8 \pi }\\\nonumber&+&\frac{3
J^2 \left(Q^2 r_+^2-2 Q^4-8 \pi Q^2 r_+^3 T+6 \alpha Q^2 w r_+^{1-3
w}+8 \pi r_+^5 T+3 \alpha  w r_+^{3-3 w}+4 r_+^4\right)}{8 \pi r_+^6
\left(2 Q^2+2 \pi r_+^3 T-\frac{3}{2} \alpha  (w+1) r_+^{1-3
w}+r_+^2\right){}^2}\\  \label{25}&+&O(J^4),
\end{eqnarray}
which is  an expansion in powers of  $J$. To avoid the complex
computation, we expand all quantities to $O(J^2)$ by neglecting all
higher order terms in $J^2$. Introducing a new variable (which
corresponds to the specific volume of the associated vdW fluid) as
\begin{eqnarray}\label{26}
\nu=2\bigg(\frac{3V}{4\pi}\bigg)^{\frac{1}{3}}=2r_++\frac{12J^2}
{r_+}\bigg(\frac{8 \pi  P r_+^4+ Q^2- \alpha  r_+^{1-3 w}+3
r_+^2}{\left(8 \pi P r_+^4+3 Q^2-3 \alpha  r_+^{1-3 w}+3
r_+^2\right){}^2}\bigg),
\end{eqnarray}
which leads Eq.(\ref{25}) to the form
\begin{eqnarray}\nonumber
P&=&\frac{T}{\nu }-\frac{1}{2 \pi  \nu ^2} +\frac{2 Q^2}{\pi  \nu
^4}-\frac{3 \alpha  2^{3 w} w \nu ^{-3 (w+1)}}{\pi }+\frac{48
J^2}{\pi  \nu ^6}\\ \nonumber &+&\frac{48 J^2}{\pi  \nu ^6 \left(\nu
^2+8 Q^2+\pi \nu ^3 T-3 \alpha 8^w (w+1) \nu ^{1-3 w}\right)^2}
\bigg(-48 q^4-12 \pi  \nu ^3 q^2 T\\ \nonumber&-&9 \alpha  q^2 2^{3
w+2} w^2 \nu ^{1-3 w}-10 \nu ^2 Q^2+5 \alpha Q^2
8^{w+1} \nu ^{1-3 w}+3 \alpha Q^2 2^{3 w+2} w \nu ^{1-3 w}\\
\nonumber&-&9 \pi \alpha T 8^w w^2 \nu ^{4-3 w}+5 \pi \alpha  T 8^w
\nu ^{4-3 w}-3 \pi \alpha T 2^{3 w+1} w \nu ^{4-3 w}\\
\nonumber&+&27 \alpha ^2 64^w w^3 \nu ^{2-6 w}+27 \alpha ^2 64^w w^2
\nu ^{2-6 w}-9 \alpha 8^w w^2 \nu ^{3-3 w}-9 \alpha ^2 64^w \nu
^{2-6 w}\\ \label{27}&-&9 \alpha ^2 64^w w \nu ^{2-6 w}+7 \alpha 8^w
\nu ^{3-3 w}+3 \alpha  8^w w \nu ^{3-3 w}\bigg).
\end{eqnarray}

The critical points can be obtained through the constraints
\begin{eqnarray}\label{28}
\frac{\partial P}{\partial \nu}|_{T=Tc}=0,\\
\label{29}\frac{\partial^2 P}{\partial \nu^2}|_{T=Tc}=0.
\end{eqnarray}
Setting $\left(\nu ^2+8 Q^2+\pi \nu ^3 T-3 \alpha 8^w (w+1) \nu
^{1-3 w}\right)=1$ and utilizing Eqs.(\ref{27}) and (\ref{28}), one
can get the critical temperature $T_c$  as
\begin{eqnarray}\nonumber
T_c&=&r^{-3 w-3} \bigg(8 q^2 \nu _c^{6 w+2}-\nu _c^{6 w+4}-9 \alpha
2^{3 w} w^2 \nu
_c^{3 w+3}-9 \alpha  2^{3 w} w \nu _c^{3 w+3}+288 J^2 \nu _c^{6 w}\\
\nonumber&-&13824 J^2 q^4 \nu _c^{6 w}-81 \alpha J^2 q^2 2^{3 w+6}
w^3 \nu
_c^{3 w+1}-27 \alpha  J^2 q^2 2^{3 w+8} w^2 \nu _c^{3 w+1}\\
\nonumber&+&75 \alpha  J^2 q^2 2^{3 w+7} \nu _c^{3 w+1}+135 \alpha
J^2 q^2 2^{3 w+6} w \nu _c^{3 w+1}-1920 J^2 q^2 \nu _c^{6 w+2}\\
\nonumber&+&243 \alpha ^2 J^2 2^{6 w+5} w^4 \nu _c^2+405 \alpha ^2
J^2 2^{6 w+5} w^3 \nu _c^2-81 \alpha  J^2 2^{3 w+4} w^3 \nu _c^{3
w+3}\\ \nonumber&+&81 \alpha ^2 J^2 2^{6 w+5} w^2 \nu _c^2-27 \alpha
J^2 2^{3 w+5} w^2 \nu _c^{3 w+3}-27 \alpha ^2 J^2 2^{6 w+6} \nu
_c^2\\ \nonumber&-&135 \alpha ^2 J^2 2^{6 w+5} w \nu _c^2+63 \alpha
J^2 2^{3 w+4} \nu _c^{3 w+3}+45 \alpha J^2 2^{3 w+5} w \nu _c^{3
w+3}\bigg)\\ \nonumber&\times&\bigg(\pi(1728 J^2 q^2 \nu _c^{3 w}+81
\alpha J^2 2^{3 w+4} w^3 \nu _c+27 \alpha  J^2 2^{3 w+6} w^2 \nu
_c\\ \label{30}&-&15 \alpha J^2 2^{3 w+5} \nu _c-9 \alpha  J^2 2^{3
w+4} w \nu _c-\nu _c^{3 w+2})\bigg)^{-1},
\end{eqnarray}
where $\nu _c$ corresponds to critical specific volume whose
expression can easily be evaluated through Eqs.(\ref{27}),
(\ref{28}) and (\ref{30}). Using Eqs. (\ref{27}) and (\ref{30}), the
critical pressure can be put in the form
\begin{eqnarray}\nonumber
P_c&=&\bigg(r^{-3 (w+1)} \nu _c^{-6 (w+1)} \bigg((96 J^2 \left(120
\left(144 J^2-1\right) Q^4+1\right) \nu _c^2 r^{3 w+3}+165888\\
\nonumber&\times& J^4 Q^2 \left(48 Q^4-1\right) r^{3 w+3}+3\ 2^{3
w+1} (16 \left(9 w \left(3 w^2+w-4\right)-16\right) J^2\\
\nonumber&+&3 w (w+1)) \alpha \nu _c^8+(-r^{3 w+3}-9\ 2^{3 w+7} J^2
Q^2 (16 (3 w (3 w (9 w+16)-10)\\
\nonumber&-&113) J^2+(-w-1) \left(9 w^2-20\right)) \alpha ) \nu
_c^6+4 (\left(192 J^2+1\right) Q^2 r^{3 w+3}-27\\
\nonumber&\times&2^{3 w+8} J^4 \left(8 (9 w (w (3 w+10)-1)-80) Q^4-3
w (3 w+2)+5\right) \alpha ) \nu _c^4) \nu _c^{9 w}\\
\nonumber&+&9 2^{6 w+5} J^2 \alpha ^2 ((64 J^2 (3 w (3 w (9 w
(3 w (w+1)-8)-76)+47)+181) Q^2\\
\nonumber&+&(16 (3 w+2) \left(9 w^2-3 w-7\right) \left(9 w^2+6
w-5\right) J^2+w (3 w (3 w (3 w+5)\\
\nonumber&+&2)-13)-3) \nu _c^2) r^{3 w+3}+9\ 2^{3 w+5} J^2 (w+1) (3
w+2) (3 w^2-1)(9 w^2\\
\nonumber&+&6 w-5) \alpha \nu _c^4) \nu _c^{3 w+2}-3\ 2^{3 w+1}
\alpha(-96 J^2 Q^2 (16 (3 w (3 w (15 w+38)-23)\\
\nonumber&-&176) J^2-3 w (3 w (w+2)-7)+10) \nu _c^2 r^{3 w+3}-768
J^4(48 (3 w (3 w (3 w\\
\nonumber&+&7)-4)-40) Q^4+3 w-9 w^2 (3 w+4)+10) r^{3 w+3}+9\ 2^{3
w+4} J^2 (w+1) \\
\nonumber&\times&\left(16 \left(9 w^2-3 w-7\right) \left(9 w^2+6
w-5\right) J^2+w (9 w (3 w+2)-11)-4\right)\\
\nonumber&\times& \alpha  \nu _c^6+(\left(w-8 J^2 \left(3 w (3
w+1)^2+4\right)\right) r^{3 w+3}+3\ 2^{3 w+10} J^4 Q^2 (3 w \\
\nonumber&\times&\left(81 w^4-378 w^2-354 w+65\right)+358) \alpha )
\nu _c^4) \nu _c^{6 w+1}-2 \left(24 J^2 Q^2-\nu _c^2\right)\\
\nonumber&\times& \left(\nu _c^4+8 \left(240 J^2-1\right) Q^2 \nu
_c^2+288 J^2 \left(48 Q^4-1\right)\right) \nu _c^{12 w+3}-81\
512^{w+1} \\
\nonumber&\times&J^4 r^{3 w+3} (w+1) (3 w+2) \left(3 w^2-1\right)
\left(9 w^2+6 w-5\right) \alpha ^3 \nu _c^3\bigg)\bigg)\bigg(2 (\pi
\nu _c^{3 w}\\
\label{31}&\times& \left(\nu _c^2-1728 J^2 Q^2\right)-3\ 2^{3 w+4}
J^2 \pi (3 w+2) \left(9 w^2+6 w-5\right) \alpha  \nu _c)\bigg)^{-1}.
\end{eqnarray}
For $w=\frac{-2}{3}$ and $Q=0$, the analytical critical points can
be computed as
\begin{eqnarray}\nonumber
\nu _c&=&2 \sqrt{3} \sqrt[4]{10} \sqrt{J},  \quad T_c=-\frac{\alpha
}{2 \pi }-\frac{12 \alpha  J^2}{\pi }+\frac{2^{3/4}}{5 \sqrt{3}
\sqrt[4]{5} \pi  \sqrt{J}},\\  \label{32} P_c&=&\frac{4\ 2^{3/4}
\sqrt{3} \alpha J^{3/2}}{\sqrt[4]{5} \pi }-\frac{720 \alpha ^2
J^4}{\pi }-\frac{27 \alpha ^2 J^2}{\pi }+\frac{1}{36 \sqrt{10} \pi
J}.
\end{eqnarray}

It is observed that for $\alpha=0$, the usual critical behavior of
Kerr BH can be recovered \cite{45}. Similarly, for $w=\frac{-2}{3}$
and $J=0$, the critical values of thermodynamic quantities for
charged AdS BH can be obtained as \cite{44}
\begin{eqnarray}\label{33}
\nu _c=2\sqrt{6}Q,  \quad T_c=\frac{1}{3\sqrt{6}\pi
Q}-\frac{\alpha}{2\pi}, \quad P_c=\frac{1}{96\pi Q^2},
\end{eqnarray}
which shows that the critical temperature as well as pressure have
the same critical behavior as that of the vdW fluid. Notice that for
$w\neq\frac{-2}{3}$, the exact critical points of charged as well as
rotating BHs cannot be computed due to lengthy expressions of the
critical specific volume. However, to observe the behavior for other
values of state parameter, one can numerically solve $\nu_c$ through
Mathematica programming. Then the critical temperature and pressure
can be obtained through Eqs.(\ref{30}) and $(\ref{31})$,
respectively.

\section{Construction of Equal-Area Law in $T-S$ Diagram}

In this section, we will study the phase diagram of quintessential
Kerr-Newman-AdS BH with the help of Maxwell equal-area law in $T-S$
differential conjugate variables. For this purpose, we consider $Q$,
$J$, $l$, $w$ and $\alpha$ as constants. The vertical axis is $T_0$
$(T_0\leq T_c)$ depending upon the horizon radius while the
horizontal axes of the two-phase coexistence region are $S_2$ and
$S_1$, respectively. In this scenario, the Maxwell equal-area law
takes the form
\begin{eqnarray} \nonumber
&T_0(S_2-S_1)=\int_{S_1}^{S_2} T dS,&
\end{eqnarray}
\begin{equation} \label{34}
=\int_{r_1}^{r_2}(\frac{4 \pi  P r^4-\frac{q^2}{2}+\frac{3}{2}
\alpha  w r^{1-3 w}+\frac{r^2}{2}}{r^2}+\frac{a^2 \left(-\frac{8}{3}
\pi  P r^4+\frac{q^2}{2}-\frac{3}{2} \alpha  w r^{1-3
w}-r^2\right)}{r^4})dr,
\end{equation}
where points $r_{2}$ and $r_1$ should satisfy
\begin{eqnarray}\label{35}
T_0&=&-\frac{a^2 \left(\frac{3}{r_ 2}-8 \pi  P r_ 2\right)-24 \pi
Pr_ 2^3+\frac{3 q^2}{r_ 2}-9 \alpha  w r_ 2^{-3 w}-3 r_ 2}{12 \pi
\left(a^2+r_ 2^2\right)},\\ \label{36} T_0&=&-\frac{a^2
\left(\frac{3}{r_ 1}-8 \pi P r_ 1\right)-24 \pi  P r_ 1^3+\frac{3
q^2}{r_ 1}-9 \alpha w r_ 1^{-3 w}-3 r_ 1}{12 \pi  \left(a^2+r_
1^2\right)}.
\end{eqnarray}
Using the above equations, one can derive
\begin{eqnarray}\nonumber
&0=8 \pi P(r_2-r_1)-q^2(\frac{1}{ r_2^3}-\frac{1}{ r_1^3})+3w\alpha
(\frac{1}{r_2^{2+3 w}}-\frac{1}{r_1^{2+3
w}})+(\frac{1}{r_2}-\frac{1}{r_1})&\\
\nonumber&+a^2\bigg(-\frac{16}{3} \pi
P(\frac{1}{r_2}-\frac{1}{r_1})+
q^2(\frac{1}{r_2^5}-\frac{1}{r_1^5})-3 \alpha  w (\frac{1}{r_2^{4+3
w}}-\frac{1}{r_1^{4+3
w}})-2(\frac{1}{r_2^3}-\frac{1}{r_1^3})\bigg),&\\\label{37}
\end{eqnarray}
and
\begin{eqnarray}\nonumber
&8 \pi T_0=+8 \pi P(r_2+r_1)-q^2(\frac{1}{ r_2^3}+\frac{1}{
r_1^3})+3w\alpha (\frac{1}{r_2^{2+3 w}}+\frac{1}{r_1^{2+3
w}})+(\frac{1}{r_2}+\frac{1}{r_1})&\\
\nonumber&+a^2\bigg(-\frac{16}{3} \pi
P(\frac{1}{r_2}+\frac{1}{r_1})+
q^2(\frac{1}{r_2^5}+\frac{1}{r_1^5})-3 \alpha  w (\frac{1}{r_2^{4+3
w}}+\frac{1}{r_1^{4+3
w}})-2(\frac{1}{r_2^3}+\frac{1}{r_1^3})\bigg),&\\ \label{38}
\end{eqnarray}
After integrating Eq.(\ref{34}), we obtain
\begin{eqnarray}\nonumber
&&\pi T_0(r_2^2-r_1^2)=\frac{r_2-r_1}{2}+\frac{4}{3} \pi P(\frac{1}{
r_2^3}-\frac{1}{ r_1^3})+\frac{q^2}{2 }(\frac{r_1-r_2}{r_1r_2})
-\frac{\alpha}{2}(\frac{1}{r_2^{3 w}}-\frac{1}{r_1^{3 w}})\\
\nonumber&&+a^2\bigg(\frac{r_1-r_2}{r_1r_2}-\frac{8}{3} \pi P
(r_2-r_1)-\frac{ q^2}{6}(\frac{1}{r_2^3}-\frac{1}{r_1^3})+\frac{3
\alpha  w }{2 (3 w+2)}(\frac{1}{r_2^{2+3 w}}-\frac{1}{r_1^{2+3
w}})\bigg).\\
\label{39}
\end{eqnarray}

Setting $x=\frac{r_1}{r_2}$ with $0\leq x \leq 1$,
Eqs.(\ref{37})-(\ref{39}) turn into
\begin{eqnarray}\nonumber
8 \pi T_0=+8 \pi P r_2(1+x)-q^2(\frac{1+x^3}{ x^3r_2^3})+3w\alpha
(\frac{1+x^{2+3 w}}{x^{2+3 w}r_2^{2+3 w}})+(\frac{1+x}{xr_2})\\
\label{40}+a^2\bigg(-\frac{16}{3} \pi P (\frac{1+x}{xr_2})+
q^2(\frac{1+x^5}{x^5r_2^5})-3 \alpha w (\frac{1+x^{4+3 w}}{x^{4+3
w}r_2^{4+3 w}})-2(\frac{1+x^3}{x^3r_2^3})\bigg),\\ \nonumber 0=8 \pi
P r_2(1-x)+q^2(\frac{1-x^3}{ x^3r_2^3}-3w\alpha
(\frac{1-x^{2+3 w}}{x^{2+3 w}r_2^{2+3 w}})-(\frac{1-x}{xr_2})\\
\label{41}+a^2\bigg(\frac{16}{3} \pi P (\frac{1-x}{xr_2})-
q^2(\frac{1-x^5}{x^5r_2^5})+3 \alpha w (\frac{1-x^{4+3 w}}{x^{4+3
w}r_2^{4+3 w}})+2(\frac{1-x^3}{x^3r_2^3})\bigg),
\end{eqnarray}
and
\begin{eqnarray}\nonumber
\pi T_0r_2^2(1-x^2)= \frac{4}{3}\pi P
r_2^3(1-x^3)-\frac{q^2}{2}(\frac{1-x}{ xr_2})+\alpha
(\frac{1-x^{3 w}}{2x^{3 w}r_2^{3 w}})+(\frac{r_2(1-x)}{2})\\
\label{42}+a^2\bigg(q^2(\frac{1-x^3}{6x^3r_2^3})-\frac{(1-x)}{r_2x}-\frac{8}{3}
\pi P r_2(1-x)-3 \alpha w (\frac{1-x^{2+3 w}}{2(2+3w)x^{2+3
w}r_2^{2+3 w}}))\bigg).
\end{eqnarray}
Through Eqs.(\ref{40}) and (\ref{42}), we get $T_0$-free relation as
\begin{eqnarray}\nonumber
&&-\frac{8 \pi  P r_2 (1-x)^2}{3 (x+1)}-\frac{q^2}{r_2^3 x}
\left(\frac{x^3+1}{x^2}-\frac{4}{x+1}\right)+\alpha r_2^{-3 w-2}
x^{-3 w} \\ \nonumber &&\times\left(\frac{3 w \left(x^{3
w+2}+1\right)}{x^2}-\frac{4 \left(1-x^{3 w}\right)}{1-x^2}\right)
+\frac{\frac{x+1}{x}-\frac{4}{x+1}}{r_2}+a^2 \bigg(-\frac{16 \pi  P
(1-x)^2}{3 r_2 x (x+1)}\\ \nonumber &&+\frac{q^2}{r_2^5 x^3}
\left(\frac{x^5+1}{x^2}-\frac{4 \left(x^2+x+1\right)}{3
(x+1)}\right)+3 \alpha  w r_2^{-3 w-4} x^{-3 w-2} \bigg(\frac{4
\left(1-x^{3 w+2}\right)}{1-x^2}\\ \label{43}&&-\frac{x^{3
w+4}+1}{x^2}\bigg)+\frac{2
\left(\frac{4}{x+1}-\frac{x^3+1}{x^2}\right)}{r_2^3 x}\bigg)=0.
\end{eqnarray}
Utilizing Eq.(\ref{41}), the explicit expression of pressure reads
\begin{eqnarray}\nonumber
8 \pi (1-x) P&=&\frac{-1}{3 r_2^6 x^5}\bigg(a^2 (-q^2 \left(-3
x^5-2 x^4+2 x+3\right)+3 \alpha  w r_2^{1-3 w} x^{1-3 w} \\
\nonumber&\times&\left(-(3 x+2) x^{3 w+3}+2 x+3\right)+2 r_2^2 x^2
\left(-3 x^3-x^2+x+3\right))\bigg)\\ \label{44}&+&\frac{q^2
\left(1-x^3\right)}{r_2^4 x^3}-\frac{1-x}{r_2^2 x}-3 \alpha w
r_2^{-3 w-3} x^{-3 w-2} \left(1-x^{3 w+2}\right).
\end{eqnarray}
Substituting the above expression in Eq.(\ref{43}) yields
\begin{eqnarray}\nonumber
&&\frac{1}{3} a^2 \bigg(- q^2 \left(x^2+x+1\right) \left(3 x^2+11
x+3\right) r_2^{-4} x^{-4}+2\left(3 x^2+13 x+3\right) r_2^{-2}
x^{-2}\\ \nonumber&&+\frac{3 \alpha  w r_2^{-3w-3}
x^{-3w-3}}{(1-x)^3}(-2 x^{3 w+3}+28 x^{3 w+4}-5 x^{3 w+5}-3 x^{3
w+6}+2 x^3-28 x^2\\ \nonumber&&+5 x+3)\bigg)+\frac{q^2 \left(x^2+4
x+1\right)}{r_2^2 x^2}-\frac{3 \alpha r_2^{-3 w-1} x^{-3
w-1}}{(1-x)^3} (w ((x+1) \left(1-x^{3 w+3}\right)\\
\label{45}&&-2 x^2 \left(1-x^{3 w}\right))-2 x^2 \left(1-x^{3
w}\right))=1,
\end{eqnarray}
which further can be written as
\begin{eqnarray} \label{46}
r_2^2=q^2f_1(x)-\frac{q^2a^2f_2(x)}{r^2}+\sigma(-f_3(x,w)r^2
+a^2f_4(x,w))+a^2f_5(x),
\end{eqnarray}
with
\begin{eqnarray}\nonumber
f_1(x)&=&\frac{x^2+4 x+1}{x^2}, \quad
f_2(x)=\frac{\left(x^2+x+1\right) \left(3 x^2+11 x+3\right)}{3
x^4},\\ \nonumber f_3(x,w)&=&\frac{x^{-3 w-1} \left(w \left((x+1)
\left(1-x^{3 w+3}\right)-2 x^2 \left(1-x^{3 w}\right)\right)-2 x^2
\left(1-x^{3 w}\right)\right)}{(1-x)^3},\\ \nonumber
f_4(x,w)&=&\frac{w x^{-3 w-3}}{3 (1-x)^3} (-2 x^{3 w+3}+28 x^{3
w+4}-5 x^{3 w+5}-3 x^{3 w+6}-2 x^3\\ \nonumber&-&28 x^2+5 x+3),
\quad f_5(x)=\frac{2 \left(3 x^2+13 x+3\right)}{3 x^2},\\ \nonumber
\sigma&=&-\frac{B_c r_2^{-3 w-1} r_c^{3 w+1}}{2 \pi  w}, \quad
B_c=-\frac{6w\pi\alpha}{r_c^{3 w+1}},
\end{eqnarray}
where $B_c$ is the quintessence of unit thickness of BH horizon.

It is observed that for $w=\frac{-2}{3}$, $f_3(x,w)$ vanishes and
when $a=0$, the results are identical with the one obtained in
charged AdS BH \cite{48}. For $x\rightarrow 1$, there must exist
$r_1=r_2=r_c$ ($r_c$ is the horizon of critical radius) which leads
Eq.(\ref{46}) to
\begin{eqnarray}\nonumber
\frac{1}{6}&=&\phi_c^2-\frac{B_c}{8\pi}(w+1)(3w+2)+\frac{4\pi
a^2}{A_c}(-\frac{17\phi_c^2}{6}+\frac{19}{9}+\frac{B_c}{54\pi}\\
\label{47}&\times&(-50+w(1+6w)(41+33w))),
\end{eqnarray}
where $\phi_c^2=\frac{q^2}{r_c^2}$ denotes the charged potential and
$A_c=4\pi r_c^2$ corresponds to the area of critical horizon. When
the values of $B_c$, $A_c$ and $w$ are given, $\phi_c$ can be
evaluated in terms of rotation parameter. Further, for given values
of $q$ and $a$, we can obtain the critical radius. Thus from
Eq.(\ref{47}), it is noted that the position of phase transition
point not only depend on the size of BH but also depends on $a$,
$B_c$, $\phi_c$ and $w$. Using $x\rightarrow 1$ in Eq.(\ref{43}),
one can obtain the critical pressure as
\begin{eqnarray}\nonumber
P_c&=&\frac{3}{8 \pi }\bigg(\frac{1}{r_c^2}-\frac{5
q^2}{r_c^4}-\frac{B (3 w+2) (3 w+4) R^{3 w+1} r_c^{-3 w-3}}{6 \pi
}+16 J^2 \bigg(-\frac{11}{r_c^4}\\ \label{48}&+&\frac{5 q^2}{3
r_c^6}-\frac{B (3 w+4) \left(18 w^2+18 w-2\right) R^{3 w+1} r_c^{-3
w-5}}{6 \pi }\bigg)\bigg),
\end{eqnarray}
and critical temperature takes the form
\begin{eqnarray}\nonumber
T_c&=&-\frac{4 q^2}{\pi r_2^3}+\frac{27 \alpha  w (w+1)^2 r_2^{-3
w-2}}{4 \pi }+\frac{1}{\pi r_2}+16 J^2 \bigg(-\frac{37}{4 \pi
r_2^3}+\frac{4 q^2}{\pi r_2^5}\\ \label{49}&+&\frac{\alpha  w
\left(\frac{81 w^3}{2}+90 w^2+\frac{81 w}{2}-\frac{43}{3}\right)
r_2^{-3 w-4}}{\pi }\bigg).
\end{eqnarray}
Using the above equations, we can derive the critical behavior of
thermodynamic quantities. It can easily be observed that this method
is more efficient than the usual approach to determine the critical
points of complicated BHs.
\begin{table}
\caption{Numerical solutions for $x$, $r_2$ and $T_0$ at constant
pressure with $J=q=0.1$ and $w=\frac{-1}{3}$.}
\begin{center}
\begin{tabular}{|c| c| c| c| c| c| c}
\hline {}$B_c$ & $\chi$ & $x$ & $r_2$ &$T_0$ \\
\hline  & 0.7 & 0.8658 &1.5853 &0.059999  \\
0.05& 0.8 & 0.8682 &1.5949 &0.06175  \\
& 0.9 & 0.87068 &1.60566 &0.06353  \\ \hline
& 0.7 & 0.9442 &5.488 &0.052899  \\
0.15& 0.8 &0.941&5.1549 &0.054042  \\
& 0.9 & 0.93934 &4.8704 &0.056097  \\ \hline & 0.7 &
0.94431&7.0782 &0.0428893\\
0.25& 0.8 &0.9421  &6.66225 &0.044072  \\
& 0.9 & 0.9397 &6.30903 &0.04678  \\ \hline
\end{tabular}
\end{center}
\end{table}

In order to find the explicit expression of $r_2$, Eq.(\ref{46}) can
be written as
\begin{eqnarray}\label{50}
r_2^2=\frac{\frac{J^2}{0.25^2} \left(-\frac{f_2(x) \left(f_3(x,w)
\sigma +1\right)}{f_1(x)}+f_4(x,w) \sigma +f_5(x)\right)}{f_3(x,w)
\sigma +1}+\frac{f_1(x) q^2}{f_3(x,w) \sigma +1}.
\end{eqnarray}
Substituting the above expression in Eq.(\ref{44}) and setting
$P_0=\chi P_c$ with $0\leq \chi\leq1$, we obtain $r_2$-free
relation. For given values of $q$, $J$, $w$, $B_c$ and $\chi$, the
numerical value of $x$ can be evaluated. Inserting the obtained
value of $x$ in Eq.(\ref{47}), we get $r_2$ and from $r_1=x r_2$, we
can have $r_1$. Finally , the value of $T_0$ (from Eq.(\ref{40}))
can be obtained. Table \textbf{1} provides the numerical values of
$x$, $r_2$ and $T_0$ for $J=0.1$, $q=0.1$ and $w=\frac{-1}{3}$. From
\begin{eqnarray}\nonumber
T&=&\frac{1}{8 \pi }(16 J^2 \left(\frac{B_c r_+^{-3 w-4} r_c^{3
w+1}}{\pi }-\frac{32 \pi  P \chi }{3 r_+}+\frac{2
q^2}{r_+^5}-\frac{4}{r_+^3}\right)-\frac{B_c r_+^{-3 w-2} r_c^{3
w+1}}{\pi }\\ \label{51} &+&16 \pi  P r_+ \chi -\frac{2
q^2}{r_+^3}+\frac{2}{r_+}),
\end{eqnarray}
we plot $T-S$ diagram for pressure $P=\chi P_c$. To analyze the
impact of parameters $w$ and $B_c$ on the phase transition point, we
plot the curves at the same pressure as shown in Figures \textbf{2}
and \textbf{3}. The horizontal black line in Figure \textbf{2}
denotes the coexistence region of two phases and its intersection
with the curve gives the position of the first-order phase
transition point. At a given pressure $P(<P_c)$, when the horizon
radius $r_+<r_1$, BH corresponds to the liquid phase of the vdW
system. For $r_+>r_2$, the BH corresponds to the vapor phase of the
vdW system. When the event horizon of BH lies in the range $r_1\leq
r_+ \leq r_2$, the BH corresponds to the vapor-liquid coexistence
phase of the vdW system.

It is observed that phase transition point and the coexistence
region of the two phases increase with increasing $B_c$ at the same
pressure. Thus, the phase transition in a charged as well as
rotating BH does not merely depend on the size of BH but electric
potential, angular momentum and state parameter also affect its
position. Moreover, the temperature of BH decreases for larger
choices of $B_c$ and $w$.  It is found that isobar in the isobaric
decreases with increasing pressure. In the coexisting regions, the
evolution of BH is different from Hawking behavior. As the
temperature of BH should be higher due to Hawking radiation but in
coexisting regions, the radiation does not affect the temperature as
well as pressure of BH.
\begin{figure}\center
\epsfig{file=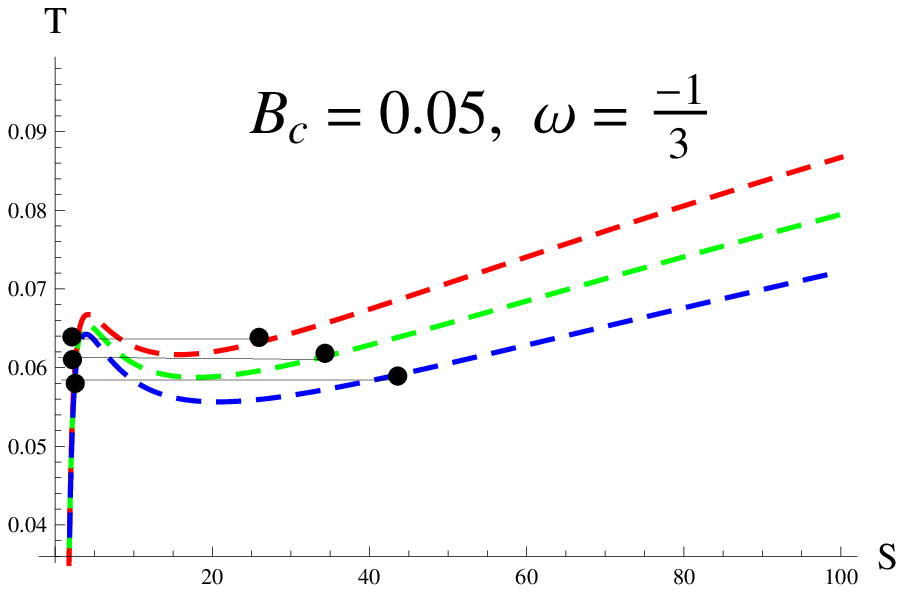,width=0.45\linewidth}
\epsfig{file=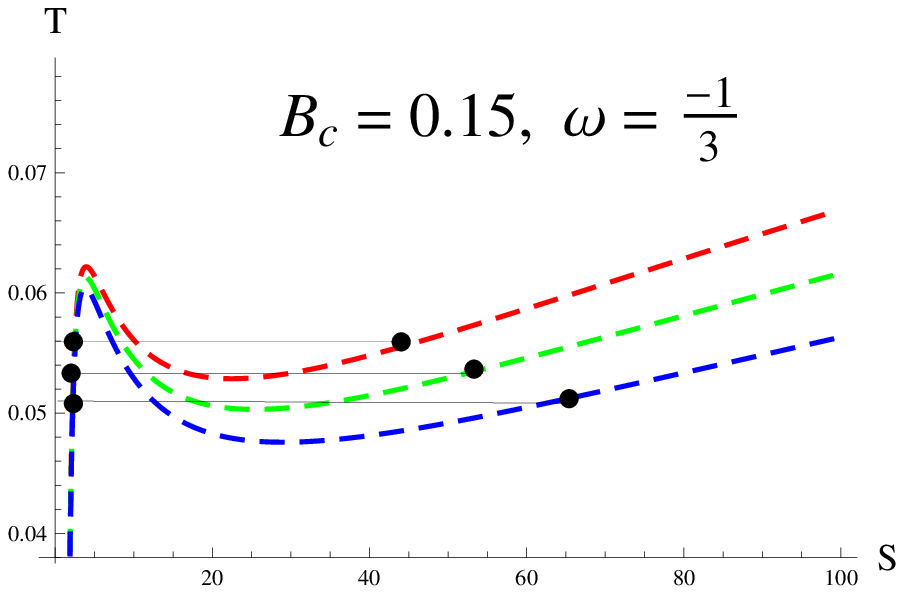,width=0.45\linewidth}
\epsfig{file=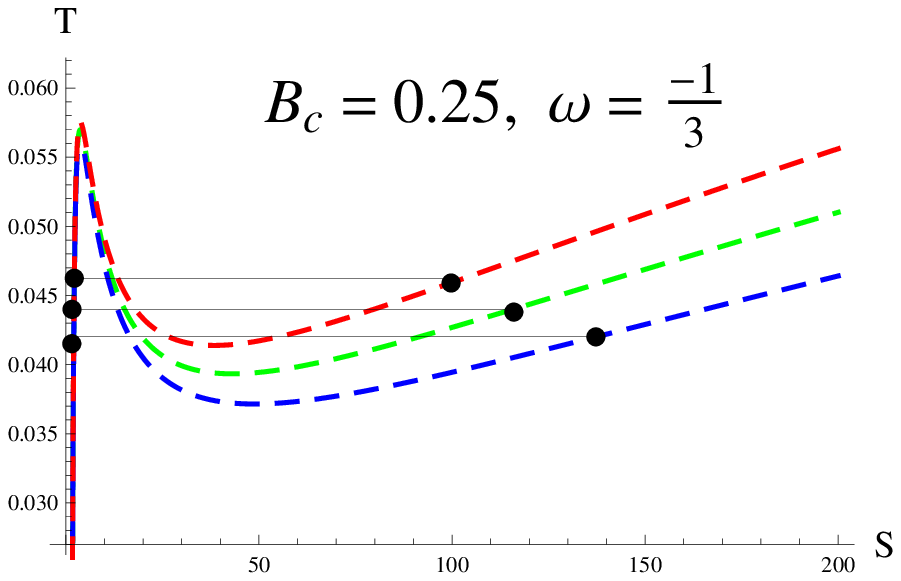,width=0.45\linewidth} \caption{Plots of
temperature vs $S$ for $J=q=0.1$ at constant pressure. The red,
green and blue curves correspond to  $\chi=0.9$, $\chi=0.8$ and
$\chi=0.7$, respectively.}
\end{figure}
\begin{figure}\center
\epsfig{file=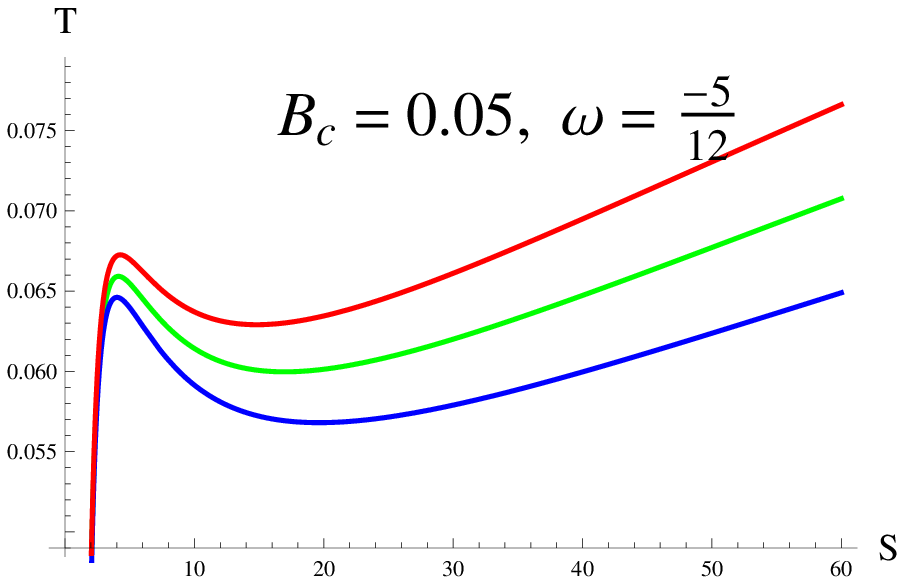,width=0.32\linewidth}
\epsfig{file=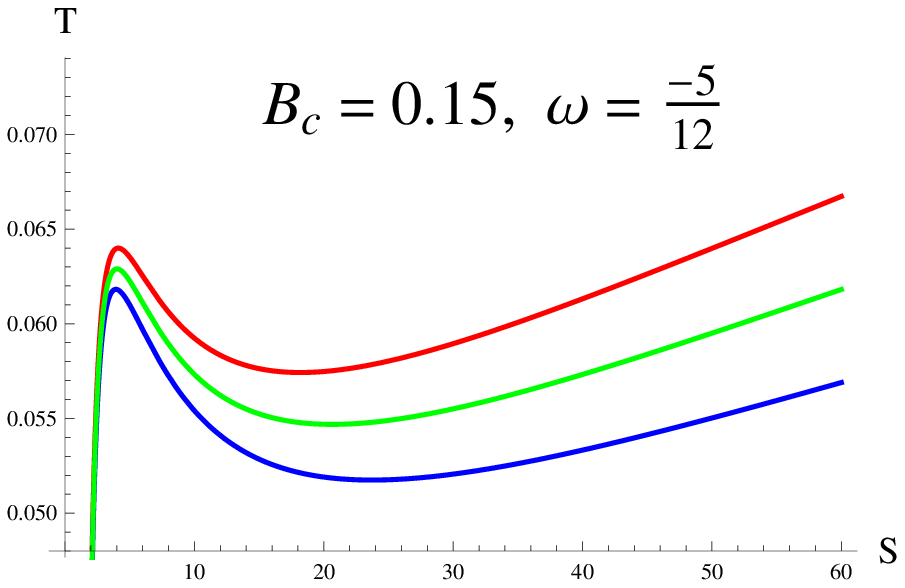,width=0.32\linewidth}
\epsfig{file=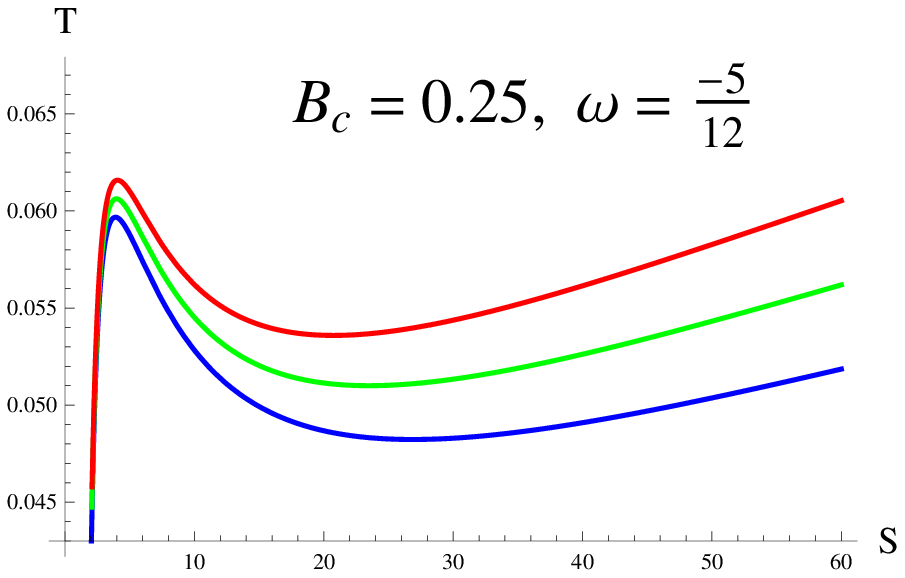,width=0.32\linewidth}\\
\epsfig{file=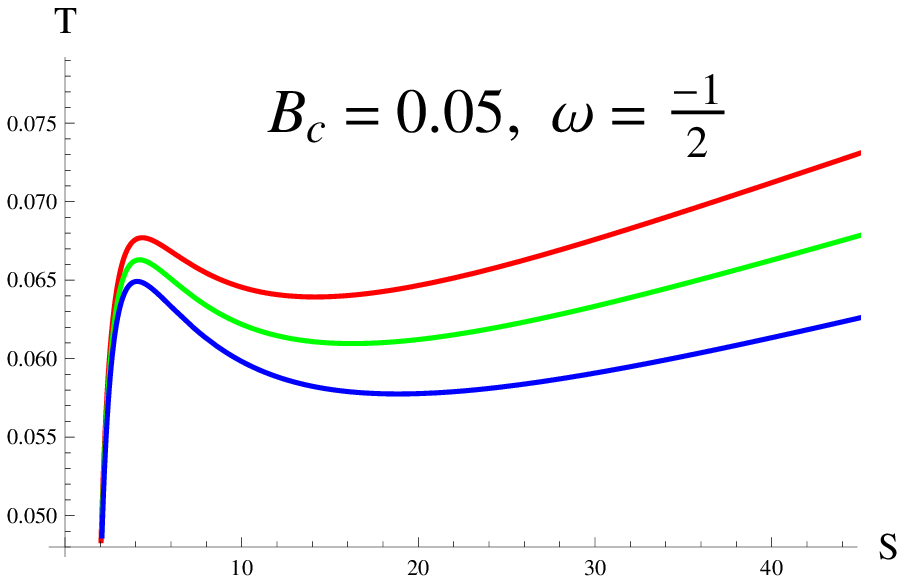,width=0.32\linewidth}
\epsfig{file=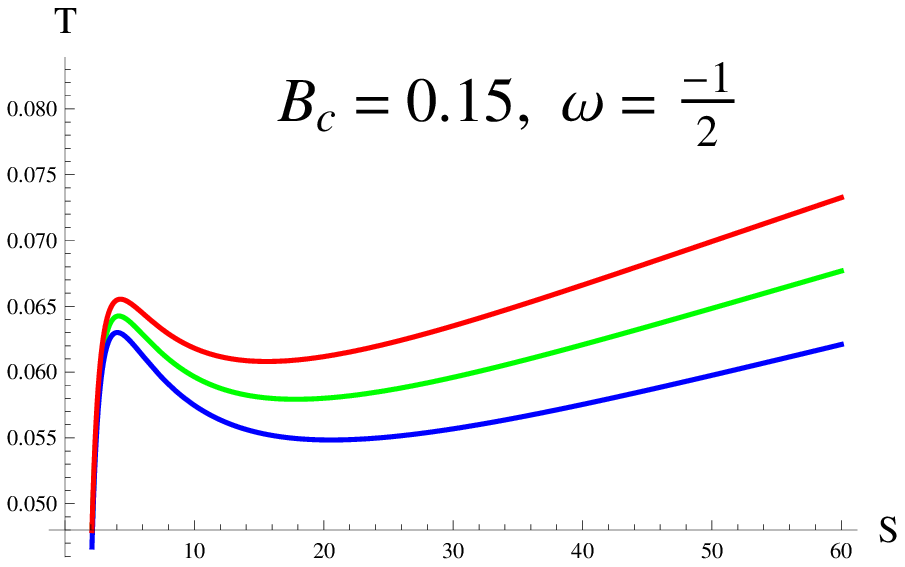,width=0.32\linewidth}
\epsfig{file=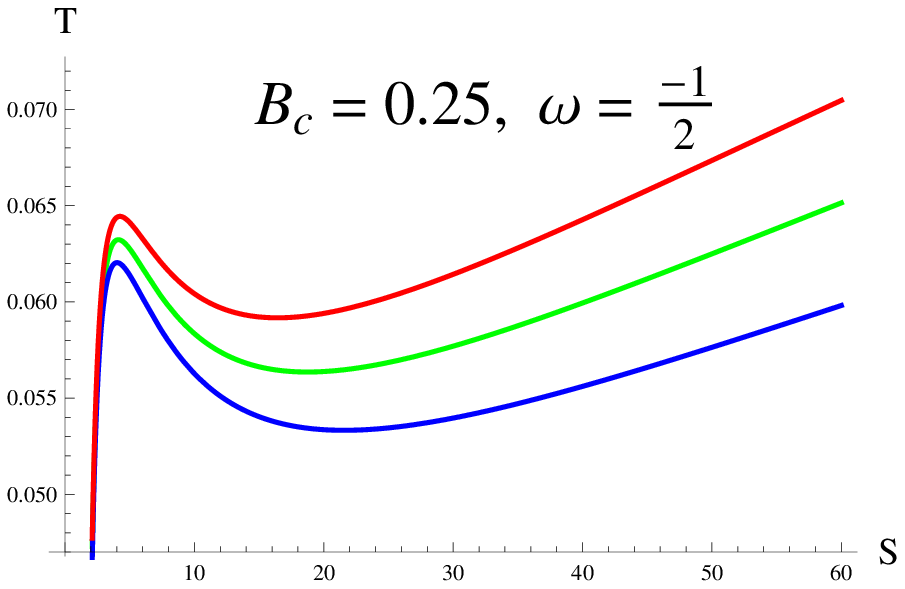,width=0.32\linewidth} \caption{Plots of
temperature vs $S$ for $J=q=0.1$ at constant pressure. The red,
green and blue curves correspond to  $\chi=0.9$, $\chi=0.8$ and
$\chi=0.7$, respectively.}
\end{figure}

\section{Thermal Fluctuations}

In this section, we investigate the effects of thermal fluctuations
on stability of Kerr-Newman-AdS BH surrounded by quintessence. We
firstly derive the exact expression of entropy in the presence of
statistical perturbations around the equilibrium state which further
yields modification in other thermodynamic potentials. To compute
the corrected entropy against thermal fluctuations, the canonical
partition function is taken to be
\begin{equation}\label{40'}
\mathcal{Z}(\beta)=\int_{0}^{\infty}\exp^{-\beta E}\sigma(E)dE,
\end{equation}
where $T=\frac{1}{\beta}$, $E$ and $\sigma(E)$ are the average
energy and quantum density of the system, respectively. Using
inverse Laplace transformation, the density of states is calculated
as
\begin{equation}\label{41'}
\sigma(E)=\frac{1}{2\pi i}\int_{b-i\infty}^{b+i\infty}\exp^{\beta
E}\mathcal{Z}(\beta)d\beta=\frac{1}{2\pi
i}\int_{b-i\infty}^{b+i\infty} \exp^{S_{0}(\beta)}d\beta,
\end{equation}
where $b>0$ and $S_{0}=\ln \mathcal{Z}+\beta E$ is the corrected
entropy. Through the method of steepest descent around saddle point
$\beta$, the above integral can be put in the form
\begin{equation}\label{42'}
S_{0}(\beta)=S+\frac{1}{2}(\beta-b)^2\frac{d^2S}{d\beta^{2}}|_{\beta=b}
+\text{higher-order terms},
\end{equation}
where $S=S_{0}(\beta)$ is the zero-order entropy with
$\frac{\partial S}{\partial\beta}=0$ and $\frac{\partial^{2}
S}{\partial\beta^{2}}>0$ at $\beta=b$. Through Eqs.(\ref{41'}) and
(\ref{42'}), we get
\begin{equation}\label{43'}
\sigma(E)=\frac{e^{S}}{2\pi i}
\int_{b-i\infty}^{b+i\infty}\exp^{\frac{1}{2}(\beta-b)^2
\frac{d^2S_{0}}{d\beta^{2}}}d\beta,
\end{equation}
which can further be simplified as
\begin{equation}\label{44'}
\sigma(E)=\frac{e^{S}}{\sqrt{2 \pi \frac{d^2S_{0}}{d\beta^{2}}}}.
\end{equation}
Eventually, this leads to
\begin{equation*}\label{45'}
S_{0}=S-\beta\ln(ST^{2})+\frac{\beta_{1}}{S},
\end{equation*}
where $\beta$ and $\beta_{1}$ are correction parameters.\\
\begin{itemize}
\item For $\beta, \beta_{1} \rightarrow 0$, the original BH entropy
(entropy without any correction parameters) can be recovered.
\item For $\beta \rightarrow 1, \beta_{1} \rightarrow 0$, the usual
logarithmic corrections can be obtained.
\item For $\beta \rightarrow 0,\beta_{1} \rightarrow 1$, the second
order correction term can be obtained which is inversely
proportional to original BH entropy.
\item Finally, for $\beta, \beta_{1} \rightarrow 1$, the effects of higher order
correction terms can be observed.
\end{itemize}
\begin{figure}\center
\epsfig{file=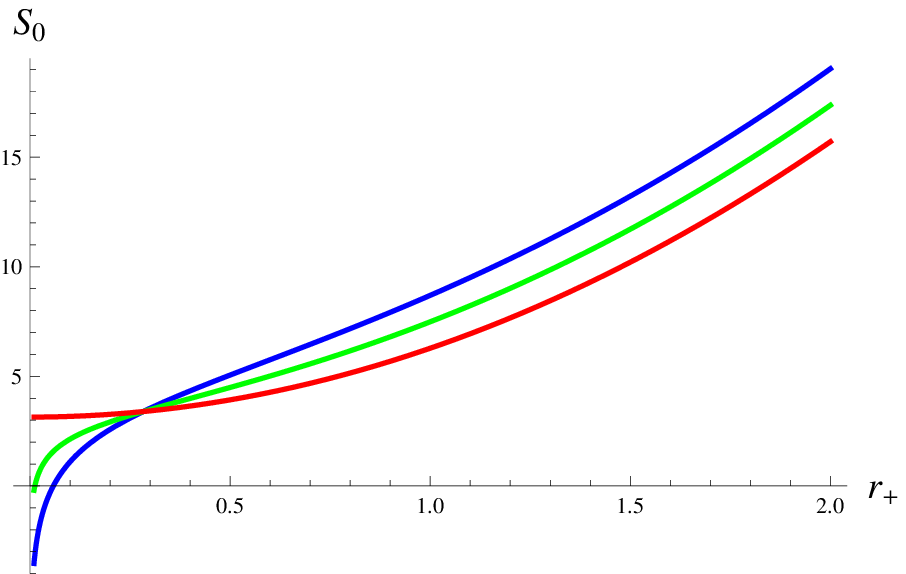,width=0.45\linewidth}
\epsfig{file=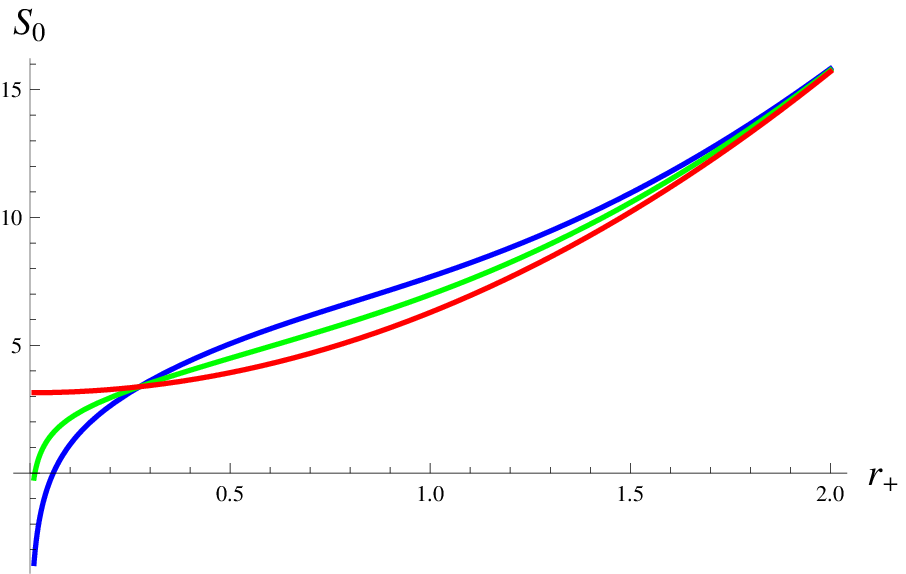,width=0.45\linewidth}\caption{Corrected entropy
vs $r_{+}$ for $a=1=q$, $\alpha=2$ with $w=\frac{-1}{3}$ (left plot)
and $w=\frac{-2}{3}$ (right plot). Here $\beta=0$, 0.5 and 1 are
denoted by red, green and blue lines, respectively.}
\end{figure}

Here, we consider the second case $(\beta \rightarrow 1,
\beta_{1}\rightarrow 0)$. It is noted that second term in the above
expression is logarithmic in nature which appears due to small
fluctuations around the equilibrium state. Since BH is regarded as a
macroscopic object while thermal fluctuations become effective on
Planck scale level, therefore the logarithmic correction terms have
a small contribution in the equilibrium entropy. Inserting
Eqs.(\ref{10}) and (\ref{11}) in the above expression, the corrected
entropy is evaluated as
\begin{eqnarray}\nonumber
S_{0}&=&\frac{\pi  \left(a^2+r_+^2\right)}{k}-\beta  \ln
\big((-\frac{\frac{8}{3} \pi  P r_+^2
\left(a^2+r_+^2\right)+a^2+q^2-\alpha  r_+^{1-3 w}+r_+^2}{r_+}\\
\nonumber&+&\frac{16}{3} \pi  a^2 P r_++\frac{32}{3} \pi  P
r_+^3+r_+^{-3 w} (-(\alpha -3 \alpha  w))+2 r_+){}^2\big)\big(16 \pi
k \left(a^2+r_+^2\right)\big)^{-1}.\\ \label{53}
\end{eqnarray}
To analyze the impact of correction terms, we plot corrected as well
as uncorrected entropy for different values of $w$ and $\alpha$.
Figures \textbf{4} and \textbf{5} show that equilibrium entropy
($\beta=0$) is positive valued as well as monotonically increasing
function which satisfies the second law of BH thermodynamics (i.e.,
entropy of BH always increases). However, in the presence of thermal
fluctuations, the entropy of small BHs becomes negative for larger
choices of correction parameter. We observe that BH gains more
entropy for larger (smaller) choices of state parameter
(quintessence parameter) which implies an increase in the area of BH
geometry. It is noted that for larger horizon radius, the corrected
entropy shows the same behavior as that of uncorrected one which
yields an important fact that thermal fluctuations do not affect the
thermodynamics of large BH.
\begin{figure}\center
\epsfig{file=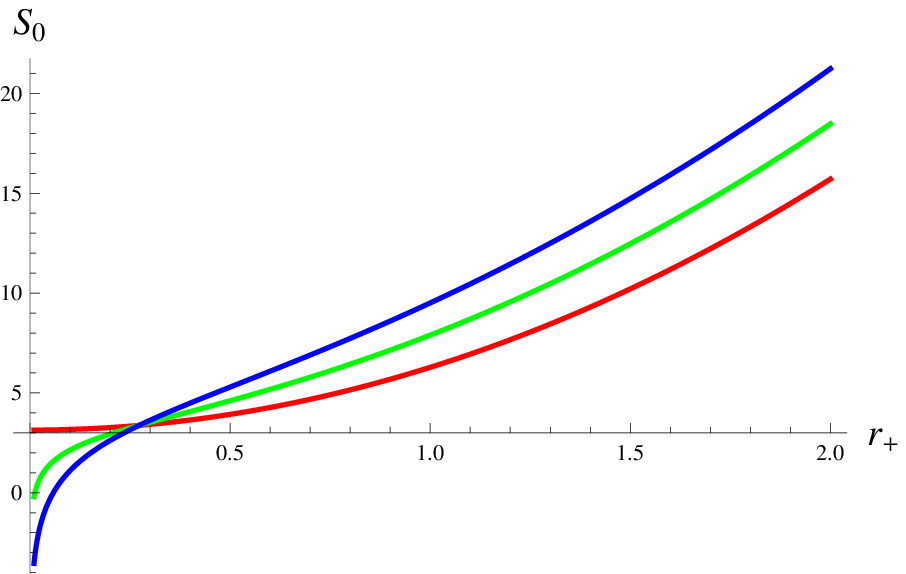,width=0.45\linewidth}
\epsfig{file=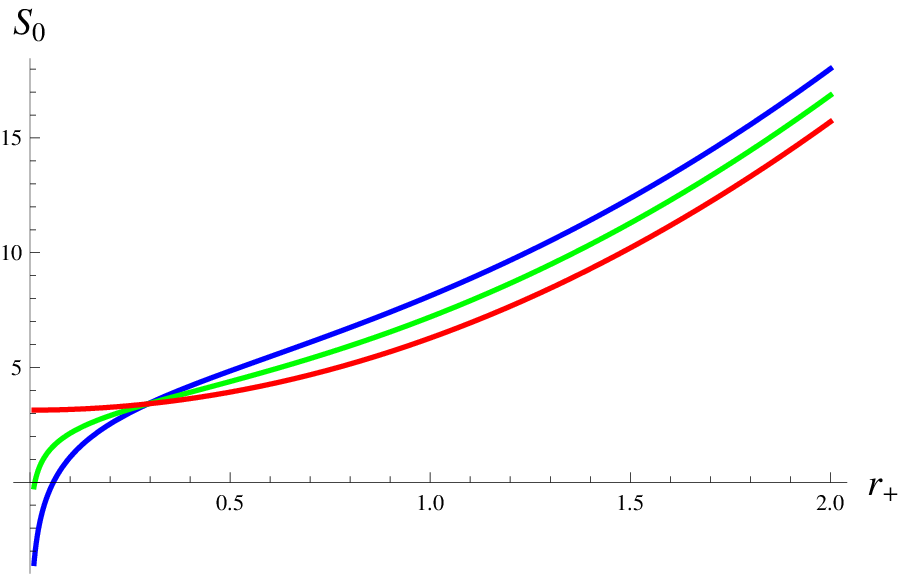,width=0.45\linewidth}\caption{Corrected entropy
vs $r_{+}$ for $a=1=q$, $w=\frac{-1}{3}$ with $\alpha=1$ (left plot)
and $\alpha=3$ (right plot). Here $\beta=0$, 0.5 and 1 are denoted
by red, green and blue lines, respectively.}
\end{figure}

The Helmholtz free energy $(F=M-TS_{0}-PV)$ is the direct measure of
work that one can get out of a system. It becomes constant once a
reversible equilibrium is achieved. The first-order corrected
Helmholtz free energy as the Legendre transformation of the internal
energy is calculated by
\begin{eqnarray}\nonumber
F&=&\frac{1}{4} r_+^{-3 w-1} \big(\big(\beta  \left(r_+^{3 w}
\left(-l^2 \left(a^2+q^2\right)+r_+^2 \left(a^2+l^2\right)+3
r_+^4\right)+3 \alpha  l^2 r_+ w\right)\\ \nonumber&\times& \ln
\left(\frac{r_+^{-6 w-2} \left(r_+^{3 w} \left(l^2
\left(a^2+q^2\right)-r_+^2 \left(a^2+l^2\right)-3 r_+^4\right)-3
\alpha  l^2 r_+ w\right){}^2}{16 \pi  l^2 \left(l^2-a^2\right)
\left(a^2+r_+^2\right)}\right)\big)\\ \nonumber&\times&\big(\pi  l^2
\left(a^2+r_+^2\right)\big)^{-1}+ \big(\alpha  l^2 r_+ \left(a^2 (3
w+1)-l^2 (3 w+2)\right)+r_+^{3 w} (l^2 \\ \nonumber&\times&\left(3
l^2-2 a^2\right) \left(a^2+q^2\right)+r_+^4 \left(4 a^2-3
l^2\right)+r_+^2 \left(2 a^4-a^2 l^2+l^4\right))\big)\\
\label{54}&\times&\big(\left(a^2-l^2\right)^2\big)^{-1}\big).
\end{eqnarray}
Figure \textbf{6} shows that when $\frac{-2}{3}< w\leq
\frac{-1}{3}$, the Helmholtz free energy attains positive values for
small BH whereas, for large BH, it shows negative behavior (left
plot). However, for $-1<w\leq\frac{-2}{3}$, the Helmholtz free
energy remains positive throughout the considered domain as shown in
the right plot of Figure \textbf{6}. It is noted that the correction
parameter increases and decreases the Helmholtz free energy before
and after the critical horizon radius, respectively.
\begin{figure}\center
\epsfig{file=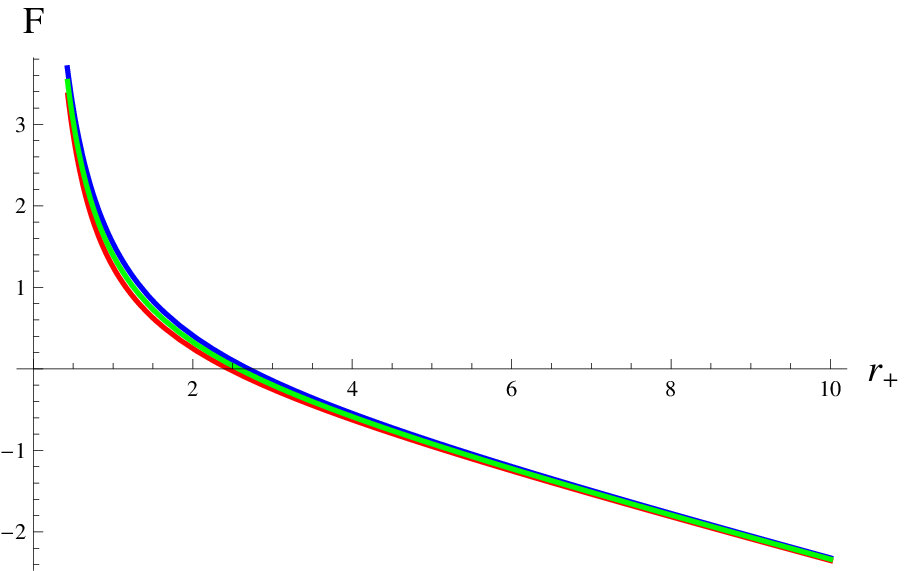,width=0.45\linewidth}
\epsfig{file=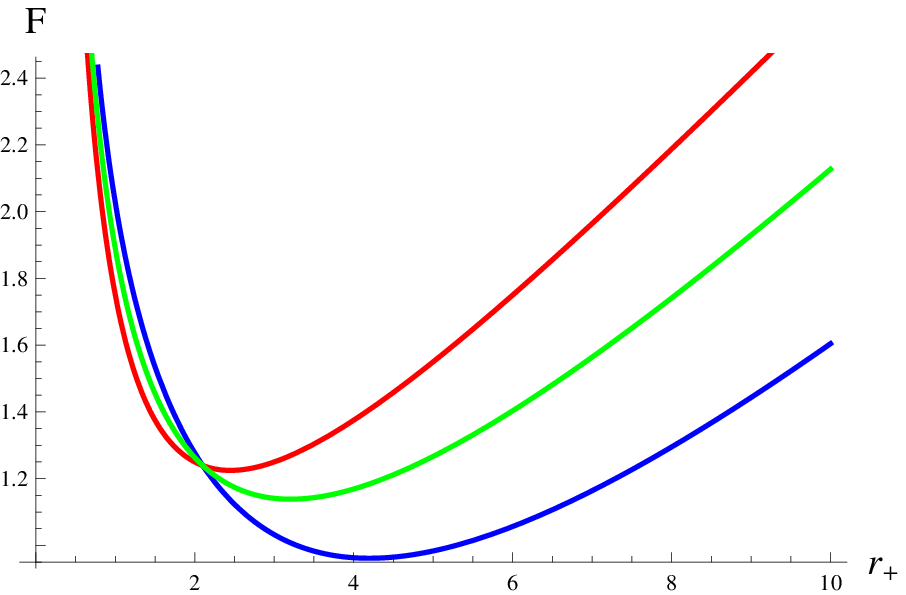,width=0.45\linewidth} \caption{Helmholtz free
energy vs $r_{+}$ for $a=1=q$, $\alpha=2$ with $w=\frac{-1}{3}$
(left plot) and $w=\frac{-2}{3}$ (right plot). Here $\beta=0$, 0.5
and 1 are denoted by red, green and blue lines, respectively.}
\end{figure}
\begin{figure}\center
\epsfig{file=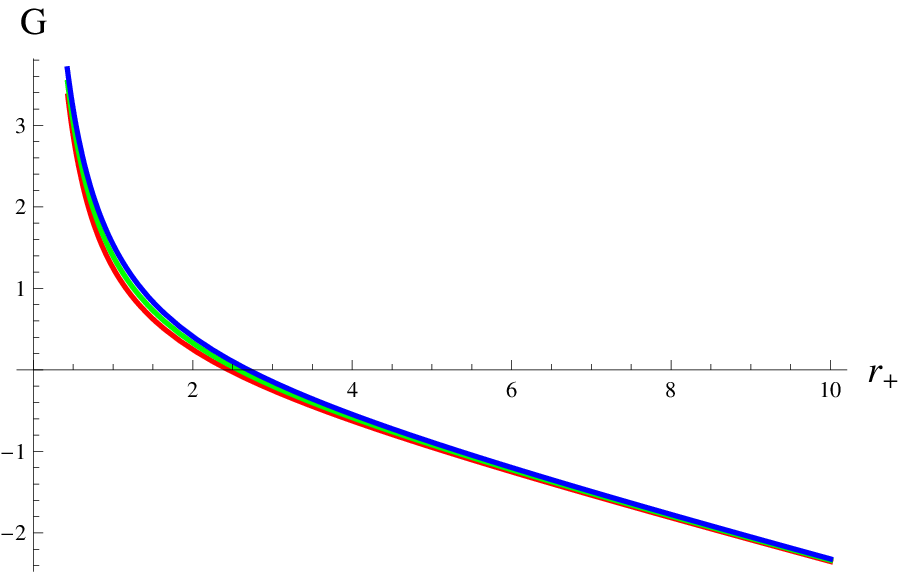,width=0.45\linewidth}
\epsfig{file=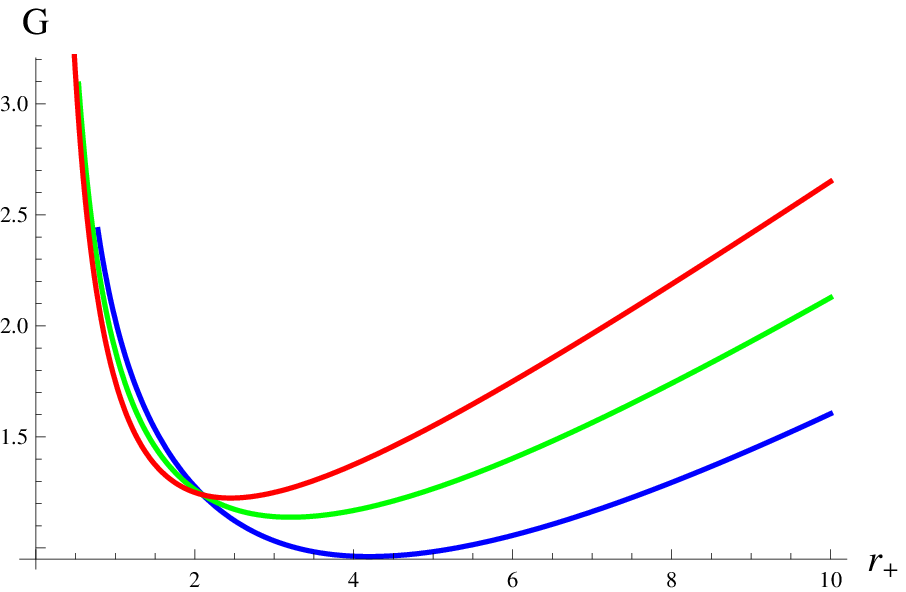,width=0.45\linewidth} \caption{Gibbs free energy
vs $r_{+}$ for $a=1=q$, $\alpha=2$ with $w=\frac{-1}{3}$ (left plot)
and $w=\frac{-2}{3}$ (right plot). Here $\beta=0$, 0.5 and 1 are
denoted by red, green and blue lines, respectively.}
\end{figure}

In the extended phase space, the mass of BH is interpreted as
enthalpy while Gibbs free energy is used to measure the maximum
amount of reversible work that may be performed by a thermodynamic
system. The corresponding Gibbs free energy $(G=M-TS_{0})$ is
derived to be
\begin{eqnarray}\nonumber
G&=&\frac{1}{4 r_+}\big(\big(r_+^{-3 w} \left(r_+^{3 w} \left(l^2
\left(a^2+q^2\right)-r_+^2 \left(a^2+l^2\right)-3 r_+^4\right)-3
\alpha  l^2 r_+ w\right)\\ \nonumber&\times& ( \ln
\left(\frac{r_+^{-6 w-2} \left(r_+^{3 w} \left(l^2
\left(a^2+q^2\right)-r_+^2 \left(a^2+l^2\right)-3 r_+^4\right)-3
\alpha  l^2 r_+ w\right){}^2}{16 \pi  l^2 \left(l^2-a^2\right)
\left(a^2+r_+^2\right)}\right)\\ \nonumber &\times& \beta
(a^2-l^2)+\pi l^2 \left(a^2+r_+^2\right))\big)\big(\pi  l^2
\left(l^2-a^2\right) \left(a^2+r_+^2\right)\big)^{-1}\\
\label{56}&+&\frac{2 \left(\frac{r_+^2
\left(a^2+r_+^2\right)}{l^2}+a^2+q^2-\alpha r_+^{1-3
w}+r_+^2\right)}{\left(\frac{a^2}{l^2}-1\right)^2}\big).
\end{eqnarray}
Figure \textbf{7} indicates that Gibbs free energy depicts the same
trend as that of Helmholtz free energy. It is known that positive
values of Gibbs energy correspond to non-spontaneous reactions that
requires an external source of energy whereas its negative values
correspond to spontaneous reactions which can be driven without any
external source. Black holes with negative Gibbs energy are
thermodynamically stable as they release their energy in the
surroundings to acquire the low energy state. It can be seen that
large BHs with $\frac{-2}{3}< w\leq\frac{-1}{3}$ (left plot) are
thermodynamically stable as $G<0$.  For $-1<w\leq\frac{-2}{3}$, the
system is unstable due to positive range of Gibbs free energy. This
indicates that larger values of state parameters yield a stable
model.

To analyze the local stability as well as phase transition, we
calculate specific heat by incorporating thermal fluctuation
effects. The divergence points of heat capacity are known as phase
transition points whereas the signature of specific heat determines
thermal stability of BH. The positive values of specific heat ensure
thermodynamical stable phase while its negative values lead the
system towards instability. The specific heat $(C=T\frac{\partial
S_{0}}{\partial T})$ is computed as
\begin{eqnarray}\nonumber
C&=&\bigg(18 \alpha  r_+ w (\beta  k \left(3 w
\left(a^2+r_+^2\right)+r_+^2\right)+\pi
r_+^2\left(a^2+r_+^2\right))+2 r_+^{3 w} (\pi  r_+^2
\\ \nonumber
&\times&\left(a^2+r_+^2\right) \left(r_+^2 \left(8 \pi  a^2
P+3\right)-3 \left(a^2+q^2\right)+24 \pi  P r_+^4\right)-\beta  k
(a^4 \\ \nonumber &\times&\left(8 \pi  P r_+^2+3\right)+3 a^2 (3
\left(8 \pi  P r_+^4+r_+^2\right)+q^2)+6 r_+^2 \left(8 \pi  P
r_+^4+q^2\right)))\bigg)\\ \nonumber &\times&\bigg(k r_+^{3 w}
(r_+^4 \left(64 \pi a^2 P-3\right)+3 a^2
\left(a^2+q^2\right)+r_+^2 \left(8 \pi  a^4 P+12 a^2+9 q^2\right)\\
\label{57} &+&24 \pi  P r_+^6)-9 \alpha  k r_+ w \left(3 a^2 w+r_+^2
(3 w+2)\right)\bigg)^{-1}.
\end{eqnarray}
From graphical analysis of Figure \textbf{8}, one can observe that
the specific heat diverges at critical radius $r_{+}=1$ which
indicates that the system undergoes the first-order phase transition
(left plot). Moreover, it is noted that the thermal fluctuations do
not affect the position of phase transition. For $\frac{-2}{3}<
w\leq\frac{-1}{3}$, BHs with large horizon radius are thermally
stable as the specific heat lies in the positive range whereas small
BHs attain its negative values which leads the system towards
instability. However, the right plot displays the negative values of
specific heat for all values of $r_+$ which indicates thermally
unstable phase of BH. It is found that larger values of correction
parameter yield more negative range of specific heat for small
horizon radius without affecting the large BH geometries. This
indicates that thermal fluctuations affect the stability of small
BHs while the large BHs mostly remain unaffected.
\begin{figure}\center
\epsfig{file=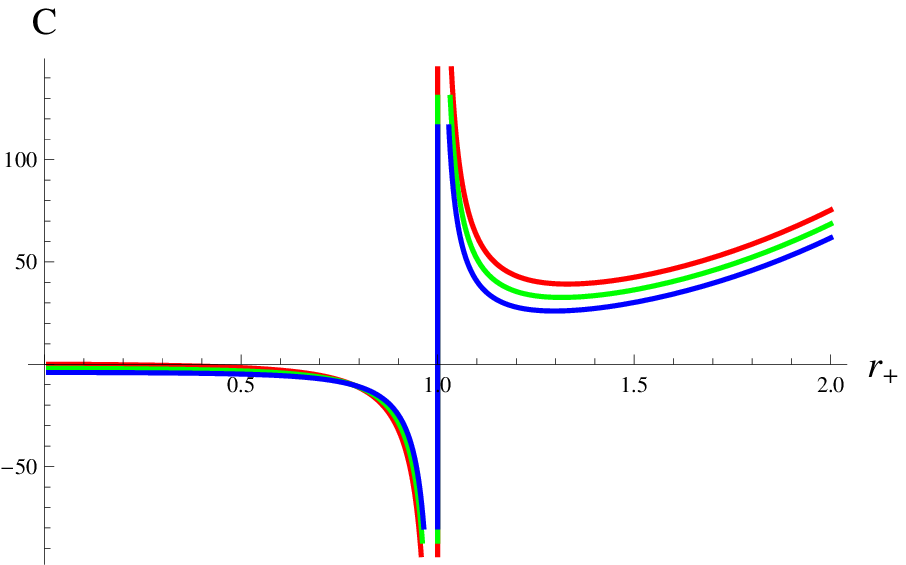,width=0.45\linewidth}
\epsfig{file=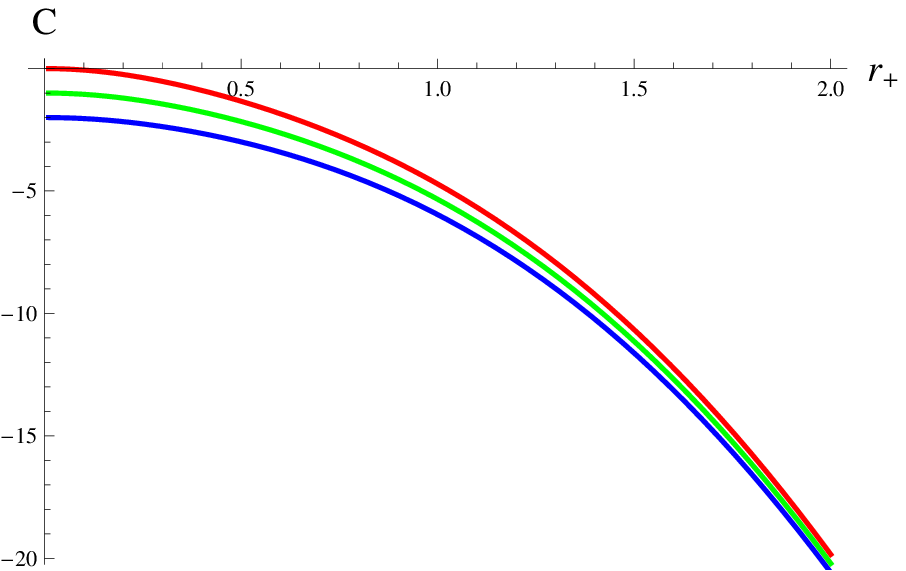,width=0.45\linewidth} \caption{specific heat vs
$r_{+}$ for $a=1=q$, $\alpha=2$ with $w=\frac{-1}{3}$ (left plot)
and $w=\frac{-2}{3}$ (right plot). Here $\beta=0$, 0.5 and 1 are
denoted by red, green and blue lines, respectively.}
\end{figure}

\section{Conclusions}

The study of critical phenomenon as well as thermal properties of
BHs has always been an interesting topic in theoretical physics. In
this paper, we have studied the effects of DE on critical behavior
and phase transition of quintessential Kerr-Newman-AdS BH. In this
regard, we have derived the exact expression of thermodynamic
quantities that satisfy Smarr-Gibbs-Dehum relation in extended phase
space. The graphical analysis of Hawking temperature shows that the
BH temperature decreases against the quintessence parameter. The
critical behavior of thermodynamic quantities are investigated
through two approaches, i.e., vdW-like equation of state and Maxwell
equal-area law. It is found that the latter technique can
efficiently discuss the critical behavior of the complicated BH.
Using equal-area law, we have also constructed the phase diagram in
$T-S$ conjugate variables and found an isobar which shows the
coexistence region of two phases. Moreover, we have computed leading
order thermal corrections to entropy to investigate the effects of
thermal fluctuations.

It is known that BH behaves like vdW liquid-gas system if isobaric
contains a region where the condition of stable equilibrium is not
satisfied. Similar to the vdW system, we have replaced un-physical
oscillating region with an isobar. Here, isobar is represented by a
black line that shows the coexistence region of two phases. Using
Maxwell equal-area law, we have found the position of isobar in
$T-S$ plane at different pressures. It is analyzed that the higher
the pressure is, the shorten of isobar will be. The temperature of
BH increases and decreases, respectively, for larger values of $B_c$
and $w$. It is noted that the coexistence region also increases for
larger choices of $B_c$. We have also observed that the considered
BH has first-order phase transition similar to that of the vdW
system which does not merely depend on the size of BH but electric
potential, angular momentum and state parameter also affect its
position.

Finally, we have analyzed the effects of thermal fluctuations by
plotting the corrected as well as uncorrected entropy for different
choices of $w$ and $\alpha$. It is interesting to mention here that
the existence of quintessence matter does not affect the uncorrected
entropy. However, the corrected entropy increases and decreases,
respectively, for larger values of $w$ and $\alpha$. The increase in
entropy BH implies an increase in the area of BH geometry. The
logarithmic corrections disturb the entropy of small BH while for
large BH, the corrected entropy coincides with the equilibrium state
indicating that statistical perturbations do not affect
thermodynamics of large BH. We have found that for $\frac{-2}{3}<
w\leq \frac{-1}{3}$, the Gibbs free energy attains positive values
for small BH whereas, for large BH, it shows negative behavior.
However, for $-1<w\leq\frac{-2}{3}$, it remains positive throughout
the considered domain which indicates that smaller values of state
parameters yield an unstable model.

To analyze the local stability, we have studied the physical
behavior of specific heat with respect to horizon radius and found
the same cut off values of $w$ as found in Gibbs energy. Thus the
model is located in thermally stable region for $\frac{-2}{3}< w\leq
\frac{-1}{3}$. It is observed that BH undergoes first-order phase
transition and its position remains unchanged under logarithmic
corrections. We conclude that thermal fluctuations yield more
unrealistic regions in small BH geometry while large BHs mostly
remain unaffected. It is worthwhile to mention here that for
$\alpha=0$, all derived quantities reduce to charged rotating AdS BH
\cite{45} and in the absence of rotation parameter, it leads to
quintessential charged AdS BH solution \cite{44,48}.

\vspace{0.5cm}

\textbf{Acknowledgement}

\vspace{0.5cm}

One of us (QM) would like to thank the Higher Education Commission,
Islamabad, Pakistan for its financial support through the
\emph{Indigenous Ph.D. Fellowship, Phase-II, Batch-III}.


\vspace{0.5cm}

\begin{thebibliography}{40}

\bibitem{25c} Stuchlik, Z.: Mod. Phys. Lett. A \textbf{20}(2005)561.

\bibitem{fa} Wei, S.W. and Liu, Y.X.: Phys. Rev. D \textbf{87}(2013)044014.

\bibitem{51} Poshteh, M.B.J., Mirza, B. and  Sherkatghanad, Z.:
Phys. Rev. D \textbf{88}(2013)024005.

\bibitem{fa2} Cai, R.G. et al.: J. High Energy Phys. \textbf{09}(2013)005.

\bibitem{fa3} Miao, Y.G. and Xu, Z.M.: Eur. Phys. J. C
\textbf{77}(2017)403.

\bibitem{b} Hendi, S.H. and Dehghani, A.:  Eur. Phys. J. C
\textbf{79}(2019)1.

\bibitem{45} Gunasekaran, S., Kubiznak, D. and Mann, R.B.: J. High Energy Phys.
\textbf{11}(2012)110.

\bibitem{52} Cheng, P., Wei, S.W. and Liu, Y.X.:
Phys. Rev. D \textbf{94}(2016)024025.

\bibitem{46} Wei, S.W. and Liu, Y.X.: Phys. Rev. D \textbf{101}(2020)104018.

\bibitem{44} Li, G.Q.: Phys. Lett. B \textbf{735}(2014)256.

\bibitem{48} Guo, X.Y. et al.: Eur. Phys. J. C \textbf{80}(2020)168.

\bibitem{1} Hawking, S.W.: Commun. Math. Phys. \textbf{43}(1975)199.

\bibitem{18} Pourhassan, B., Faizal, M. and Debnath, U.: Eur. Phys. J. C
\textbf{76}(2016)145; Pourhassan, B., Kokabi, K. and Rangyan, S.:
Gen. Relativ. Gravit. \textbf{49}(2017)144; Pourhassan, B., Kokabi,
K. and Sabery, Z.: Ann. Phys. \textbf{399}(2018)181.

\bibitem{41'} Upadhyay, S.: Gen. Relativ. Gravit.
\textbf{50}(2018)128.

\bibitem{21} Sinha, A.K.: arxiv:1608.08359; Upadhyay, S. et al.: Prog. Theor. Exp. Phys.
\textbf{2018}(2018)093E01.

\bibitem{22a} Zhang, M.: Nucl. Phys. B \textbf{935}(2018)170.

\bibitem{23a} Sharif, M. and Akhtar, Z.: Phys. Dark Universe \textbf{29}(2020)100589.

\bibitem{25f} Newman, E. and Penrose, R.: Phys. Rev. Lett.
\textbf{11}(1962)566.

\bibitem{26a} Xu, Z. and Wang, J.: Phys. Rev. D
\textbf{95}(2017)064015.

\bibitem{1a} Bekenstein, J.D.: Phys. Rev. D \textbf{7}(1973)2333; Hawking, S.W.: Commun. Math. Phys. \textbf{46}(1976)206.

\bibitem{3a} Bekenstein, J.D.: Lett. Nuovo Cimento
\textbf{4}(1972)737.

\bibitem{4a} Bardeen, J.M., Carter, B. and Hawking, S.W.: Commun. Math. Phys. \textbf{31}(1973)161.

\bibitem{5a} Hawking, S.W.: Nature \textbf{248}(1974)30; Bekenstein, J.D.: Phys. Rev. D \textbf{9}(1974)3292.

\bibitem{qa} Gibbons, G.W., Perry, M.J. and Pope, C.N.: Class. Quantum Grav. \textbf{22}(2005)1503.

\bibitem{ma} Anabalon, A., et al.: J. High Energy Phys. \textbf{2019}(2019)96.

\bibitem{a} Kubiznak, D. and Mann. R.B.: J. High Energy Phys. \textbf{7}(2012)33.

\bibitem{mm} Li, G.Q.: Phys. Lett. B \textbf{735}(2014)256.

\end{thebibliography}
\end{document}